\documentclass[twocolumn]{aastex63}

\usepackage{natbib}
\usepackage{amsmath,amssymb}
\usepackage{xspace}
\usepackage{rotating}
\usepackage{multirow}
\usepackage{dcolumn}
\usepackage{comment}
\usepackage[caption=false]{subfig}	

\definecolor{mpl_blue}{HTML}{1F77B4}
\definecolor{mpl_orange}{HTML}{FF7F0E}
\definecolor{mpl_green}{HTML}{2CA02C}
\definecolor{mpl_red}{HTML}{D62728}

\newcommand{\kink}{``kink''}
\renewcommand{\fig}[1]{Fig.~\ref{#1}}
\newcommand{\eqn}[1]{Eqn.~\ref{#1}}

\newcommand{\sect}[1]{\S~\ref{#1}}

\usepackage[all]{hypcap}
\usepackage[caption=false]{subfig}
\shorttitle{The NANOGrav 11-Year Data Set: Evolution of Gravitational Wave Background Statistics}
\shortauthors{The NANOGrav Collaboration}

\defcitealias{dfg+13}{NG5a}
\defcitealias{abb+14}{NG5b}
\defcitealias{abb+15}{NG9a}
\defcitealias{abb+17}{NG11a}
\defcitealias{abb+17b}{NG11b}

\graphicspath{{./}{figures/}}
\begin{document}

\title{The NANOGrav 11-Year Data Set: Evolution of Gravitational Wave Background Statistics}

%%%%%%%%%%%AUTHORS%%%%%%%%%%%%%%
\correspondingauthor{J.~S.~Hazboun}
\email{hazboun@uw.edu}
%The NANOGrav Collaboration:
\author[0000-0003-2742-3321]{J.~S.~Hazboun}
\altaffiliation{NANOGrav Physics Frontiers Center Postdoctoral Fellow}
\affiliation{University of Washington Bothell, 18115 Campus Way NE, Bothell, WA 98011, USA}

\author[0000-0003-1407-6607]{J.~Simon}
\affiliation{Jet Propulsion Laboratory, California Institute of Technology, 4800 Oak Grove Drive, Pasadena, CA 91109, USA}
\affiliation{Theoretical AstroPhysics Including Relativity (TAPIR), MC 350-17, California Institute of Technology, Pasadena, California 91125, USA}

\author{S.~R.~Taylor}
\affiliation{Department of Physics and Astronomy, Vanderbilt University, 2301 Vanderbilt Place, Nashville, TN 37235, USA}
\affiliation{Jet Propulsion Laboratory, California Institute of Technology, 4800 Oak Grove Drive, Pasadena, CA 91109, USA}
\affiliation{Theoretical AstroPhysics Including Relativity (TAPIR), MC 350-17, California Institute of Technology, Pasadena, California 91125, USA}

\author[0000-0003-0721-651X]{M.~T.~Lam}
\affiliation{School of Physics and Astronomy, Rochester Institute of Technology, Rochester, NY 14623, USA}
\affiliation{Department of Physics and Astronomy, West Virginia University, P.O.~Box 6315, Morgantown, WV 26506, USA}
\affiliation{Center for Gravitational Waves and Cosmology, West Virginia University, Chestnut Ridge Research Building, Morgantown, WV 26505, USA}

\author[0000-0003-4700-9072]{S.~J.~Vigeland}
\affiliation{Center for Gravitation, Cosmology and Astrophysics, Department of Physics, University of Wisconsin-Milwaukee,\\ P.O.~Box 413, Milwaukee, WI 53201, USA}

\author{K.~Islo}
\affiliation{Center for Gravitation, Cosmology and Astrophysics, Department of Physics, University of Wisconsin-Milwaukee,\\ P.O.~Box 413, Milwaukee, WI 53201, USA}

\author[0000-0003-0123-7600]{J.~S.~Key}
\affiliation{University of Washington Bothell, 18115 Campus Way NE, Bothell, WA 98011, USA}
%K.~Aggarwal\altaffilmark{6,7}, %6,7

\author{Z.~Arzoumanian}
\affiliation{X-Ray Astrophysics Laboratory, NASA Goddard Space Flight Center, Code 662, Greenbelt, MD 20771, USA}

\author[0000-0003-2745-753X]{P.~T.~Baker}
\affiliation{Department of Physics and Astronomy, Widener University, One University Place, Chester, PA 19013, USA}
\affiliation{Department of Physics and Astronomy, West Virginia University, P.O.~Box 6315, Morgantown, WV 26506, USA}
\affiliation{Center for Gravitational Waves and Cosmology, West Virginia University, Chestnut Ridge Research Building, Morgantown, WV 26505, USA}

\author{A.~Brazier}
\affiliation{Department of Astronomy, Cornell University, Ithaca, NY 14853, USA}
\affiliation{Cornell Center for Advanced Computing, Ithaca, NY 14853, USA}

%M.~R.~Brinson\altaffilmark{6},

\author{P.~R.~Brook}
\affiliation{Department of Physics and Astronomy, West Virginia University, P.O.~Box 6315, Morgantown, WV 26506, USA}
\affiliation{Center for Gravitational Waves and Cosmology, West Virginia University, Chestnut Ridge Research Building, Morgantown, WV 26505, USA}

\author[0000-0003-4052-7838]{S.~Burke-Spolaor}
\affiliation{Department of Physics and Astronomy, West Virginia University, P.O.~Box 6315, Morgantown, WV 26506, USA}
\affiliation{Center for Gravitational Waves and Cosmology, West Virginia University, Chestnut Ridge Research Building, Morgantown, WV 26505, USA}

%S.~J.~Chamberlin\altaffilmark{5}, 

\author[0000-0002-2878-1502]{S.~Chatterjee}
\affiliation{Department of Astronomy, Cornell University, Ithaca, NY 14853, USA}

%B.~Christy\altaffilmark{6},

\author[0000-0002-4049-1882]{J.~M.~Cordes}
\affiliation{Department of Astronomy, Cornell University, Ithaca, NY 14853, USA}

\author[0000-0002-7435-0869]{N.~J.~Cornish}
\affiliation{Department of Physics, Montana State University, Bozeman, MT 59717, USA}

\author[0000-0002-2578-0360]{F.~Crawford}
\affiliation{Department of Physics and Astronomy, Franklin \& Marshall College, P.O.~Box 3003, Lancaster, PA 17604, USA}

\author{K.~Crowter}
\affiliation{Department of Physics and Astronomy, University of British Columbia, 6224 Agricultural Road, Vancouver, BC V6T 1Z1, Canada}

\author[0000-0002-6039-692X]{H.~T.~Cromartie}
\affiliation{University of Virginia, Department of Astronomy, P.O.~Box 400325, Charlottesville, VA 22904, USA}

\author[0000-0002-2185-1790]{M.~DeCesar}
\altaffiliation{NANOGrav Physics Frontiers Center Postdoctoral Fellow}
\affiliation{$^{17}$Department of Physics, Lafayette College, Easton, PA 18042, USA}

\author{P.~B.~Demorest}
\affiliation{National Radio Astronomy Observatory, 1003 Lopezville Rd., Socorro, NM 87801, USA}

\author[0000-0001-8885-6388]{T.~Dolch}
\affiliation{Department of Physics, Hillsdale College, 33 E.~College Street, Hillsdale, Michigan 49242, USA}

\author{J.~A.~Ellis}
\affiliation{Infinia ML, 202 Rigsbee Avenue, Durham, NC 27701, USA}
\affiliation{Department of Physics and Astronomy, West Virginia University, P.O.~Box 6315, Morgantown, WV 26506, USA}
\affiliation{Center for Gravitational Waves and Cosmology, West Virginia University, Chestnut Ridge Research Building, Morgantown, WV 26505, USA}

\author[0000-0002-2223-1235]{R.~D.~Ferdman}
\affiliation{Department of Physics, University of East Anglia, Norwich, UK}

\author{E.~Ferrara}
\affiliation{NASA Goddard Space Flight Center, Greenbelt, MD 20771, USA}

\author{E.~Fonseca}
\affiliation{Department of Physics, McGill University, 3600  University St., Montreal, QC H3A 2T8, Canada}

\author{N.~Garver-Daniels}
\affiliation{Department of Physics and Astronomy, West Virginia University, P.O.~Box 6315, Morgantown, WV 26506, USA}
\affiliation{Center for Gravitational Waves and Cosmology, West Virginia University, Chestnut Ridge Research Building, Morgantown, WV 26505, USA}

\author[0000-0001-8158-683X]{P.~Gentile}
\affiliation{Department of Physics and Astronomy, West Virginia University, P.O.~Box 6315, Morgantown, WV 26506, USA}
\affiliation{Center for Gravitational Waves and Cosmology, West Virginia University, Chestnut Ridge Research Building, Morgantown, WV 26505, USA}

\author{D.~Good}
\affiliation{Department of Physics and Astronomy, University of British Columbia, 6224 Agricultural Road, Vancouver, BC V6T 1Z1, Canada}

%R.~Haas\altaffilmark{16},

\author[0000-0003-4143-8132]{A.~M.~Holgado}
\affiliation{NCSA and Department of Astronomy, University of Illinois at Urbana-Champaign, Urbana, Illinois 61801, USA}

\author{E.~A.~Huerta}
\affiliation{NCSA and Department of Astronomy, University of Illinois at Urbana-Champaign, Urbana, Illinois 61801, USA}

\author{R.~Jennings}
\affiliation{Department of Astronomy, Cornell University, Ithaca, NY 14853, USA}

\author{G.~Jones}
\affiliation{Department of Physics, Columbia University, New York, NY 10027, USA}

\author[0000-0001-6607-3710]{M.~L.~Jones}
\affiliation{Department of Physics and Astronomy, West Virginia University, P.O.~Box 6315, Morgantown, WV 26506, USA}
\affiliation{Center for Gravitational Waves and Cosmology, West Virginia University, Chestnut Ridge Research Building, Morgantown, WV 26505, USA}

\author{A.~R.~Kaiser}
\affiliation{Department of Physics and Astronomy, West Virginia University, P.O.~Box 6315, Morgantown, WV 26506, USA}
\affiliation{Center for Gravitational Waves and Cosmology, West Virginia University, Chestnut Ridge Research Building, Morgantown, WV 26505, USA}

\author[0000-0001-6295-2881]{D.~L.~Kaplan}
\affiliation{Center for Gravitation, Cosmology and Astrophysics, Department of Physics, University of Wisconsin-Milwaukee,\\ P.O.~Box 413, Milwaukee, WI 53201, USA}

%V.~M.~Kaspi\altaffilmark{16},

\author{L.~Z.~Kelley}
\affiliation{Center for Interdisciplinary Exploration and Research in Astrophysics (CIERA), Northwestern University, Evanston, IL 60208}

\author{T.~J.~W.~Lazio}
\affiliation{Jet Propulsion Laboratory, California Institute of Technology, 4800 Oak Grove Drive, Pasadena, CA 91109, USA}

\author[0000-0002-2034-2986]{L.~Levin}
\affiliation{Jodrell Bank Centre for Astrophysics, University of Manchester, Manchester, M13 9PL, United Kingdom}

\author[0000-0003-4137-7536]{A.~N.~Lommen}
\affiliation{Haverford College, 370 Lancaster Ave, Haverford, PA 19041, USA}

\author[0000-0003-1301-966X]{D.~R.~Lorimer}
\affiliation{Department of Physics and Astronomy, West Virginia University, P.O.~Box 6315, Morgantown, WV 26506, USA}
\affiliation{Center for Gravitational Waves and Cosmology, West Virginia University, Chestnut Ridge Research Building, Morgantown, WV 26505, USA}

\author{J.~Luo}
\affiliation{University of Texas at San Antonio, San Antonio, TX, USA}
\affiliation{Center for Advanced Radio Astronomy, University of Texas Rio Grande Valley, Brownsville, TX 78520, USA}

\author[0000-0001-5229-7430]{R.~S.~Lynch}
\affiliation{Green Bank Observatory, P.O.~Box 2, Green Bank, WV 24944, USA}

\author[0000-0003-2285-0404]{D.~R.~Madison}
\affiliation{Department of Physics and Astronomy, West Virginia University, P.O.~Box 6315, Morgantown, WV 26506, USA}
\affiliation{Center for Gravitational Waves and Cosmology, West Virginia University, Chestnut Ridge Research Building, Morgantown, WV 26505, USA}

%C.~D.~McGrath\altaffilmark{6},

\author[0000-0001-7697-7422]{M.~A.~McLaughlin}
\affiliation{Department of Physics and Astronomy, West Virginia University, P.O.~Box 6315, Morgantown, WV 26506, USA}
\affiliation{Center for Gravitational Waves and Cosmology, West Virginia University, Chestnut Ridge Research Building, Morgantown, WV 26505, USA}

\author{S.~T.~McWilliams}
\affiliation{Department of Physics and Astronomy, West Virginia University, P.O.~Box 6315, Morgantown, WV 26506, USA}
\affiliation{Center for Gravitational Waves and Cosmology, West Virginia University, Chestnut Ridge Research Building, Morgantown, WV 26505, USA}

\author[0000-0002-4307-1322]{C.~M.~F.~Mingarelli}
\affiliation{Center for Computational Astrophysics, Flatiron Institute, 162 Fifth Avenue, New York, NY 10010, USA}

\author{C.~Ng}
\affiliation{Dunlap Institute for Astronomy and Astrophysics, University of Toronto, 50 St. George Street, Toronto, ON M5S 3H4, Canada}

\author[0000-0002-6709-2566]{D.~J.~Nice}
\affiliation{Department of Physics, Lafayette College, Easton, PA 18042, USA}

\author[0000-0001-5465-2889]{T.~T.~Pennucci}
\affiliation{Hungarian Academy of Sciences MTA-ELTE ``Extragalatic Astrophysics Research Group'', Institute of Physics, \\E\"{o}tv\"{o}s Lor\'{a}nd University, P\'{a}zm\'{a}ny P. s. 1/A, 1117 Budapest, Hungary}

\author[0000-0002-8826-1285]{N.~S.~Pol}
\affiliation{Department of Physics and Astronomy, West Virginia University, P.O.~Box 6315, Morgantown, WV 26506, USA}
\affiliation{Center for Gravitational Waves and Cosmology, West Virginia University, Chestnut Ridge Research Building, Morgantown, WV 26505, USA}

\author[0000-0001-5799-9714]{S.~M.~Ransom}
\affiliation{University of Virginia, Department of Astronomy, P.O.~Box 400325, Charlottesville, VA 22904, USA}
\affiliation{National Radio Astronomy Observatory, 520 Edgemont Road, Charlottesville, VA 22903, USA}

\author[0000-0002-5297-5278]{P.~S.~Ray}
\affiliation{Naval Research Laboratory, Washington DC 20375, USA}

\author{X.~Siemens}
\affiliation{Center for Gravitation, Cosmology and Astrophysics, Department of Physics, University of Wisconsin-Milwaukee,\\ P.O.~Box 413, Milwaukee, WI 53201, USA}

\author[0000-0002-6730-3298]{R.~Spiewak}
\affiliation{Centre for Astrophysics and Supercomputing, Swinburne University of Technology, PO Box 218, Hawthorn, VIC 3122, Australia}

\author[0000-0001-9784-8670]{I.~H.~Stairs}
\affiliation{Department of Physics and Astronomy, University of British Columbia, 6224 Agricultural Road, Vancouver, BC V6T 1Z1, Canada}

\author[0000-0002-1797-3277]{D.~R.~Stinebring}
\affiliation{Department of Physics and Astronomy, Oberlin College, Oberlin, OH 44074, USA}

\author[0000-0002-7261-594X]{K.~Stovall}
\affiliation{National Radio Astronomy Observatory, 1003 Lopezville Rd., Socorro, NM 87801, USA}

\author[0000-0002-1075-3837]{J.~Swiggum}
\altaffiliation{NANOGrav Physics Frontiers Center Postdoctoral Fellow}
\affiliation{Center for Gravitation, Cosmology and Astrophysics, Department of Physics, University of Wisconsin-Milwaukee,\\ P.O.~Box 413, Milwaukee, WI 53201, USA}

\author[0000-0002-2451-7288]{J.~E.~Turner}
\affiliation{Department of Physics and Astronomy, West Virginia University, P.O.~Box 6315, Morgantown, WV 26506, USA}
\affiliation{Center for Gravitational Waves and Cosmology, West Virginia University, Chestnut Ridge Research Building, Morgantown, WV 26505, USA}
\affiliation{Center for Gravitation, Cosmology and Astrophysics, Department of Physics, University of Wisconsin-Milwaukee,\\ P.O.~Box 413, Milwaukee, WI 53201, USA}

\author[0000-0002-4162-0033]{M.~Vallisneri}
\affiliation{Jet Propulsion Laboratory, California Institute of Technology, 4800 Oak Grove Drive, Pasadena, CA 91109, USA}
\affiliation{Theoretical AstroPhysics Including Relativity (TAPIR), MC 350-17, California Institute of Technology, Pasadena, California 91125, USA}

\author[0000-0002-6428-2620]{R.~van~Haasteren}
\altaffiliation{Currently employed at Microsoft Corporation}
\affiliation{Jet Propulsion Laboratory, California Institute of Technology, 4800 Oak Grove Drive, Pasadena, CA 91109, USA}

\author[0000-0002-6020-9274]{C.~A.~Witt}
\affiliation{Department of Physics and Astronomy, West Virginia University, P.O.~Box 6315, Morgantown, WV 26506, USA}
\affiliation{Center for Gravitational Waves and Cosmology, West Virginia University, Chestnut Ridge Research Building, Morgantown, WV 26505, USA}

\author[0000-0001-5105-4058]{W.~W.~Zhu}
\affiliation{CAS Key Laboratory of FAST, Chinese Academy of Science, Beijing 100101, China}

\collaboration{63}{(The NANOGrav Collaboration)}

%%%%%%%%%%%%ABSTRACT%%%%%%%%%%%%%%%%%%%
\begin{abstract}
An ensemble of inspiraling supermassive black hole binaries should produce a stochastic background of very low frequency gravitational waves. This stochastic background is predicted to be a power law, with a spectral index of -2/3, and it should be detectable by a network of precisely timed millisecond pulsars, widely distributed on the sky. This paper reports a new ``time slicing'' analysis of the 11-year data release from the North American Nanohertz Observatory for Gravitational Waves (NANOGrav) using 34 millisecond pulsars. Methods to flag potential ``false positive'' signatures are developed, including techniques to identify responsible pulsars. Mitigation strategies are then presented. We demonstrate how an incorrect noise model can lead to spurious signals, and show how independently modeling noise across 30 Fourier components, spanning NANOGrav's frequency range, effectively diagnoses and absorbs the excess power in gravitational-wave searches. This results in a nominal, and expected, progression of our gravitational-wave statistics. Additionally we show that the first interstellar medium event in PSR J1713+0747 pollutes the common red noise process with low-spectral index noise, and use a tailored noise model to remove these effects.
\end{abstract}

\keywords{
Gravitational waves --
Methods:~data analysis --
Pulsars:~general
}

%%%%%%%%%%%%INTRODUCTION%%%%%%%%%%%%%%%%%%%
\section{Introduction}
\label{sec:intro}
Pulsar timing arrays \citep[PTAs;][]{saz78, det79, fb90} are poised to detect the stochastic background of gravitational waves (GWs) from a population of super-massive binary black holes (SMBBHs) within approximately the next five years \citep{2013CQGra..30v4015S,rsg15,2016ApJ...819L...6T,2017MNRAS.471.4508K}. There are three PTA collaborations that have been in operation for over a decade: 
the North American Observatory for Gravitational Waves (NANOGrav; \citealt{ml13}), 
the European Pulsar Timing Array (EPTA; \citealt{dcl+16}), 
and the Parkes Pulsar Timing Array (PPTA; \citealt{h13}). A number of emerging collaborations, including the Chinese Pulsar Timing Array \citep{2016ASPC..502...19L}, the Indian Pulsar Timing Array \citep{2018JApA...39...51J} and telescope-centered timing groups such as MeerTime \citep{2018arXiv180307424B} and CHIME/Pulsar \citep{2018IAUS..337..179N}, all have a component of their programs directed toward nanohertz GW detection and characterization.
Together with the more established PTAs these groups form the International Pulsar Timing Array Collaboration (IPTA; \citealt{v+16})

The NANOGrav collaboration has so far released four data sets based on, respectively, five years of precision pulsar-timing observations (\citealt{dfg+13}; hereafter \citetalias{dfg+13}), nine years of observations (\citealt{abb+15}; hereafter \citetalias{abb+15}), 11 years of observations (\citealt{abb+17}; hereafter \citetalias{abb+17}), and 12.5 years of observations (Alam et al. in prep\footnote{The release is available at \url{https://data.nanograv.org}}). The present analysis was carried out on \citealt{abb+17}, since the newest data release has only recently been available.

The dominant signal expected at nanohertz GW frequencies (where the regime of sensitivity is set by the cadence $\Delta t$ and total baseline $T_\mathrm{total}$ of the pulsar time-series sampling: $1/T_\mathrm{total} < f < 1/2\Delta t$) is the stochastic background of GWs from a SMBBH population \citep[see e.g.,][]{2018arXiv181108826B}. 
There are several models in the literature that predict the amplitude and spectral shape of this gravitational wave background (GWB) \citep[e.g.,][]{s13,mop14,ss16,2017MNRAS.471.4508K}. 
These models employ a range of galaxy surveys, galaxy evolution scenarios, and simulations to identify the most likely demographics of SMBBHs detectable by PTAs. Recent results from NANOGrav (\citealt{abb+17b}; hereafter \citetalias{abb+17b}) and other PTAs \citep{ltm+15,srl+15} have reported constraints on the GWB characteristic strain amplitude that intersect astrophysically interesting regions of SMBBH parameter space. 
Using techniques developed in \citet{scm15}, \citet{ss16}, and \citet{tss17}, the 11-year data set was used to constrain the relationship of super-massive black hole masses to that of their host galaxies, as well as galactic center environments that may influence the final parsec of binary dynamical evolution.

Most searches for the GWB rightly focus on the most recent data set, first searching for, and then, in the absence of a signal, setting upper limits (ULs) on, the GWB. 
These results are often juxtaposed with earlier work from shorter data sets to illustrate the gains in sensitivity of these galactic-scale GW detectors. 
Here we analyze the past evolution of our statistics by slicing the \citetalias{abb+17} data set in time, and performing the analyses from \citetalias{abb+17b} on each slice. 
This allows us to characterize the growth of NANOGrav's GW sensitivity as a function of time, as well as diagnose previously unmodeled noise processes.
With regards to the latter, in this article we discuss how a noise transient produced a high-significance false-positive during the time span between the release of \citetalias{dfg+13} and \citetalias{abb+15}.
Understanding this spurious signal, and tracking down the pulsars from which it originates, will be the subject of most of this paper. 

The paper is organized as follows.  In \S~\ref{sec:obs} we discuss our methods for obtaining the pulsar timing data used in this study, and in \S~\ref{sec:data_analysis} we review our data analysis methods, including a detailed introduction to our noise models. In \S~\ref{sec:motivation} we discuss the motivation for understanding the evolution of our GW statistics and introduce theoretical models for the evolution of that signal. In \S~\ref{sec:initial_slice_results} we present the results of the initial time-slice analysis, including anomalous evidence for GWs in out data set. We then turn, in \S~\ref{sec:results}, to identifying which pulsars are responsible for this behavior and elucidate the various data analysis methods, noise model and other mitigation strategies used to understand and remove it. In \S~\ref{sec:lowspec} we connect the shallow spectral index of the common process in early slice analyses to the first interstellar medium (ISM) event in PSR J1713+0747 and show how models extant in the literature can mitigate this behavior. And finally in \S~\ref{sec:conclusions} we conclude with a summary of the issues encountered in this analysis and possible paths forward for future PTA noise mitigation.

%%%%%%%%%% THE 11-YEAR DATA SET%%%%%%%%%%%%%%%%%%%%
\section{The $11$-year Data Set}
\label{sec:obs}
The NANOGrav 11-year data set contains observations of 45 pulsars made between 2004 and 2015. 
Details of the observations and pulsars can be found in \citetalias{abb+17}. 
We briefly describe the data set here.

We made observations using two radio telescopes: 
the 100-m Robert C.~Byrd Green Bank Telescope (GBT) 
of the Green Bank Observatory in Green Bank, West Virginia; 
and the 305-m William E. Gordon Telescope (Arecibo) 
of Arecibo Observatory in Arecibo, Puerto Rico. 
Since Arecibo is more sensitive than the GBT, 
all pulsars that can be observed from Arecibo 
($0^\circ < \delta < 39^\circ$) were observed with it, 
while those outside Arecibo's declination range 
were observed with the GBT. 
Two pulsars were observed with both telescopes: 
PSRs J1713+0747 and B1937+21. 
We observed most pulsars once a month. 
In addition, we started a high-cadence observing campaign in 2013, 
in which we made weekly observations 
of two pulsars with the GBT (PSRs J1713+0747 and J1909$-$3744) 
and five pulsars with Arecibo (PSRs J0030+0451, J1640+2224, J1713+0747, J2043+1711, and J2317+1439). 

At the GBT, the monthly observations used the 820 MHz and 1.4 GHz receivers, 
while weekly observations used only the 1.4 GHz receiver. 
At Arecibo, pulsars were observed with two of four possible receivers 
(327 MHz, 430 MHz, 1.4 GHz, and 2.3 GHz), though always including the 1.4 GHz receiver. 
Backend instrumentation was upgraded about midway through our project 
from the ASP and GASP systems, which had bandwidths of 64 MHz, 
to the wideband systems PUPPI and GUPPI processing up to 800 MHz for certain receivers \citep{drd+08}. 

For each pulsar, the observed times of arrival (TOAs) were fit to a timing model 
that described the pulsar's spin period and spin period derivative, 
sky location, proper motion, and distance. To this model were added a number of parameters that describe the radio-frequency dependent behavior of the pulse arrival times.
Additionally, for those pulsars in binaries 
the timing model we also included five Keplerian parameters that described the binary orbit, 
and additional post-Keplerian parameters that described relativistic binary effects 
if they statistically improved the timing fit. 

%%%%%%%%%%%%%DATA ANALYSIS METHODS%%%%%%%%%%%%%%%%%%
\section{Data Analysis Methods}
\label{sec:data_analysis}
The analysis techniques in this work largely follow the stochastic signal procedures in \citetalias{abb+17b}. The JPL solar system ephemeris DE436 \citep{de436} was used along with the TT(BIPM2016) timescale. 
Our model likelihood is based on pulsar timing residuals, constructed for each pulsar $\delta \mathbf{t}$ as
\begin{equation}
	\delta \mathbf{t} = T \mathbf{b} + \mathbf{n} + \mathbf{s} \,.
\end{equation}
$T \mathbf{b}$ describes noise contributions modeled with Gaussian processes, including uncertainties in the pulsar timing model and low-frequency time-correlated (red) noise, $\mathbf{n}$ describes white noise (WN), and $\mathbf{s}$ describes residuals induced by a GWB, also modeled with a Gaussian process. 

The WN is modeled using the rms template-fitting errors for the TOAs. These are inflated using additional pieces, one added in quadrature (EQUAD), and a multiplicative factor (EFAC),
\begin{equation}
	\sigma^{2}_{\rm total} = \mathrm{EFAC}^{2}\sigma_{\rm TOA}^2+\mathrm{EQUAD}^2\;.
\end{equation}
In practice we build the WN correlation matrix by adding these diagonal contributions to the off-diagonal pieces, that model the correlated WN between TOAs observed in different sub bands during the same observation (ECORR) \citep{abb+14}.

The standard likelihood for gravitational-wave analysis with PTAs is well documented in the literature \citep{lah+14,lha+13,Lentati:2016ygu,ltm+15,srl+15,vanHaasteren:2012hj,vanHaasteren:2014faa,dfg+13,abb+16,tlb+17a,abb+17b}. Here we focus on a more detailed introduction to the types of time-correlated (red) noise models we use for the analyses. 

\subsection{Red Noise Models and the GWB}

The precision TOAs of radio pulses from millisecond pulsars have been used to measure myriad astrophysical interactions. Perhaps most famous is the observation of a negative binary period derivative accurately explained by the emission of gravitational waves in the context of general relativity \citep{Taylor:1982zz}. Pulsar timing measurements are also responsible for the first detection of an extrasolar planet \citep{Wolszczan:1992zg} and are used to monitor the content and movement of the galactic ionized interstellar medium \citep[e.g.,][]{Keith:2013,Jones:2017}. In fact, observations from many pulsars have been put together to map the interstellar medium (ISM) content of the galaxy \citep{Cordes:2002wz, ymw17}. Lensing events from the ISM can also be monitored using pulsars, and a recent re-occurrence of an apparent lensing event in PSR J1713+0747 has been studied extensively in \cite{Lam:2017duo}. 

From the perspective of a search for gravitational waves in pulsar timing data, these astrophysical interactions are considered sources of noise, i.e., they must be removed or mitigated in order to detect a GW. The deterministic processes are modeled by a pulsar timing model, however, in order to account for the stochastic astrophysical signals various types of models are used in our GW search analyses \citep{Lentati:2016ygu, Lam:2017duo, Madison:2019zot}. 

Since the GWB manifests as a low-frequency, time-correlated stochastic process (a ``red'' spectrum) it is especially important to model astrophysical noise sources that leave a similar signature in our timing residuals. In our analyses both the GWB and the red noise intrinsic to a pulsar's line of sight are built with the same types of models. Most commonly they are built using a normal-kernel Gaussian process in a Fourier basis with a power law prior \citep{rw06,vhv14,Lentati:2016ygu},
\begin{equation}
P=\frac{A_{\rm GWB}^2}{12\pi^2}\left(\frac{f}{f_{yr}}\right)^{-\gamma}\;{\rm yr}^3 \; ,
\end{equation}
as a power spectral density.
The spectral index parameter $\gamma$ prior is restricted from $0$ to $7$, meaning that the model must be either ``red'' (higher power at lower frequencies) or ``flat''(``white''), i.e., $\gamma=0$. The signal spectrum is then built using a Fourier basis from $N$ frequencies (often 30 in our analyses). The prior used in the Gaussian process is an ansatz for the type of time-correlated process that one expects to find in the residuals and describes the power spectral density of that stochastic process modeled in the frequency domain. The same frequency domain describes the GWB and non-GW red noise sources, and spans the nanohertz regime. 

Using a power law is the simplest model, however a few more complex models have been used in the literature, including a turnover model, a free spectral model, and trained Gaussian process models \citep{lah+2013, tgl13}. 
The free spectral model is the most generic model for a time-correlated stochastic process. This allows for a different coefficient for the Fourier basis at each frequency and is not restricted by any model for the power spectral density. While this model is very flexible, it incurs a large Occam penalty since it involves a large parameter volume.
As in \cite{abb+16} and \cite{abb+17b} the models used to search for the GWB and mitigate noise in individual pulsar data sets are not dependent on the radio frequency of the TOAs, hence we refer to these as ``achromatic'' red-noise models\footnote{This nomenclature is sometimes confusing as there are two frequency domains, the frequencies of the GWB and red noise and the {\it radio} frequencies of the pulsar observations.}.

The flagship Bayesian analysis for a PTA gravitational-wave search includes the Hellings-Downs (HD) spatial correlations \citep{hd83}. In practice these analyses are often the final analysis completed since the non-diagonal correlation matrix inversion is computationally expensive in the Bayesian framework.
The HD correlation {\it Bayesian search} takes advantage of the spatial correlations between pulsars and the time correlations due to the GWB, modeled as an achromatic red noise Gaussian process. In the weak signal regime, the autocorrelations within pulsars are a reasonable first estimate for a correlated stochastic background and have the same spectral content. In much of this manuscript we discuss this latter type of search for a common red noise process, since we will need to run numerous iterations of search types over the 18 time slices we have made.

As mentioned above, the models used for the GWB and other time-correlated processes particular to the pulsar lines-of-sight are very similar. Historically, the {\it usual} model for red noise intrinsic to a pulsar, and its line-of-sight, is modeled with a power law with varying amplitude and spectral index ($\gamma$). In the most common GW analyses, the common noise process caused by the GWB is also modeled as a power law, however, the spectral index is set to that expected for a GWB, $\gamma=13/3$. However, as we will see, it is also informative to allow the spectral index to vary when searching for the GWB. Effectively, these models are identical, but for each pulsar the model for intrinsic red noise has its own set of parameters and no spatial correlations between other pulsars are considered. Unlike the power law model, the free spectral model allows one to analyze noise independently at multiple frequencies, and as we will see, this can help to disentangle degeneracy between the noise process unique to the pulsar and the GWB. Additionally, the other functional forms of the spectral models mentioned above can be used for both pulsar red noise and the GWB.

``Chromatic'' (radio-frequency-dependent) versions of the noise models, most often modeling dispersion with a $1/\nu^2$-dependence on radio frequency, can be found in the literature \citep{Lee:2014sza,Reardon:2016,caballeroetal:2016,Lentati:2016ygu}, and were recently used in \cite{Lam:2017duo} regarding the previously mentioned ISM events in the timing of PSR~J1713+0747. The standard NANOGrav analysis has not included these types of noise models, instead using a piece-wise binning of dispersion measure (DM; the integrated line-of-sight electron density causing the $1/\nu^2$-dependence in the arrival times) fluctuations, called DMX, implemented as part of the timing model using the pulsar-timing software {\tt TEMPO}/{\tt TEMPO2} \citep{abb+17}. This method works well at describing broadband dispersion measure fluctuations that have a timescale longer than individual DMX bins \citep[$\le 1$~week;][]{Jones:2017}.  There are a significant number of possible chromatic effects \citep[e.g.,][]{cs10,Lam:2017ysu}, primarily due to radio propagation through the ISM, which need to be modeled appropriately \citep{sc17}. The total chromatic noise assuming misestimation of DM was performed by \citet{Lam:2016iie} on \citetalias{abb+15}.

\subsection{Slicing the Data Set}
Here we use the methods of \citetalias{abb+17b} to analyze various slices of the NANOGrav 11-year data set presented in \citetalias{abb+17}.
The data set was partitioned by setting Modified Julian Day (MJD) cutoffs in six-month increments after an initial three year span. 
Three years is the nominal length of individual pulsar data sets used in \citetalias{abb+17b}, and hence, adopted here as the minimal time span of data needed to do a worthwhile analysis.
The slices were cumulative, adding 6 months of data at a time. 
In order to understand the noise evolution in the pulsars, we performed single-pulsar noise analyses at every slice, where all of the white noise and red noise parameters are allowed to vary in a Bayesian analysis.
This follows from the general philosophy throughout this investigation \textemdash\ use the information known at the time of each slice to do the analysis.
The WN maximum likelihood values from these analyses were then used to set the WN parameters for the full PTA analyses, analogous to \citetalias{abb+17b}.

The Bayesian analysis was done using the {\tt enterprise} software suite \citep{enterprise} and the Markov Chain Monte Carlo (MCMC) sampling software {\tt PTMCMCSampler} \citep{evh17b}. Detection statistics were acquired by using log-uniform priors on the red noise amplitudes for both the individual pulsar red noise and the common red noise process. ULs were acquired by running analyses using linear exponential priors (meant to emulate a uniform prior but sampled in log space) for the amplitudes.

A frequentist analysis was also undertaken using the same software above and the optimal statistic submodule in the PTA model software package {\tt enterprise\_extensions} \citep{enterprise_ext}. A noise-marginalized analysis \citep{viet17} was done at each slice using the MCMC chains from the Bayesian runs to sample over the red noise parameters. The maximum likelihood values were then used to calculate the noise-maximized values. In both cases the optimal statistic and signal-to-noise ratio (SNR) were calculated.

In most cases the spectral index for the common red noise process was set to $\gamma=13/3$, the theoretical spectral index (in terms of timing residuals) for a stochastic GWB originating from binary inspirals, where the loss of energy in the binary is driven by the radiation of gravitational waves \citep{Phinney:2001di}. In addition, an analysis was done where the common process's spectral index was also allowed to vary.

%%%%%%%%%%%%%EVOLUTION OF GWB STATISTICS%%%%%%%%%%%%%%%
\section{Evolution of GWB Statistics}
\label{sec:motivation}
The first signal detected by PTAs is expected to be the stochastic sum of SMBBHs from the cosmological neighborhood \citep{rsg15} and should grow very steeply as our data sets become sensitive further into the nanohertz regime. Unlike the first detections of GWs from compact binary coalescences \citep{aaa+16, aaa+17}, a detection of the GWB will not appear as a single event, but rather a steady growth in significance over the course of a number of data releases. The evolution of detection statistics has been studied in the literature extensively \citep{sejr13,Vigeland:2016nmm},   including theoretical studies of the scaling of the frequentist optimal statistic and numerical simulations using realistic data to predict when PTAs will reach specified sensitivities. Work of this kind is important for understanding the context of current data releases and the near-future ability to characterize nanohertz GW astrophysics. These types of studies also have obvious applications to the strategic planning of future PTA facilities. 

\subsection{Theory} \label{subsec:evolution_theory}

The scaling laws presented in \citet{sejr13} provide a straightforward framework for the comparison of NANOGrav's evolving sensitivity to the GWB. In the simple case of a PTA where all pulsars only have (identical) white noise (WN), the expectation value of the signal-to-noise ratio (SNR), $\rho=A^2_{\rm GWB}/\sigma_0$, where $\sigma_0$ is the standard deviation of $A^2_{\rm GWB}$, is shown to evolve in the weak signal regime as 
\begin{equation}\label{eq:weak_scaling}
\left<\rho_{\rm weak} \right> = \left(\sum_I \sum_{J>I}\chi^2_{IJ}\right)^{\frac{1}{2}}A^2_{\rm GWB}\frac{bcT^\gamma}{\sigma^2 \sqrt{4\gamma-2}} \;,
\end{equation}
where $\chi_{IJ}$ is the HD spatial correlation between pulsars $I$ and $J$, $c$ is the cadence of observations, $\sigma$ is the measurement error of TOAs, $\gamma$ is the spectral index of the power law background, $T$ is the total time of observations, $A_{\rm GWB}$ is the amplitude of the GWB at $1/{\rm year}$ ($f_{\rm yr}$), and $b$ contains the frequency dependence of the GWB signal,
\begin{equation}
b = \frac{1}{24\pi^2}\left(\frac{1}{f_{\rm ref}}\right)^{\gamma+3} \; .
\end{equation}
Similarly in the intermediate signal regime the SNR scales as
\begin{equation}\label{eq:intermediate_scaling}
\left<\rho_{\rm int}\right> = \left(\sum_I \sum_{J>I}\chi^2_{IJ}\right)^{\frac{1}{2}} \left[2\theta\left(A^2_{\rm GWB}\frac{bc}{2\sigma^2}\right)^{\frac{1}{\gamma}}T\right]^{\frac{1}{2}} \; ,
\end{equation}
where $\theta$ is a function of the spectral index that includes the $\Gamma$ function,
\begin{equation}
\theta(\gamma) = \frac{\gamma-1}{\gamma}\Gamma\left(1-\gamma^{-1}\right)\Gamma\left(1+\gamma^{-1}\right) \; . 
\end{equation}

The SNR can be related to the Bayes factor, $\mathcal{B}_{10}$, from a GWB model versus noise-only model comparison using the Laplace approximation \citep{MacKay:2002:ITI:971143,Romano:2016dpx},
\begin{equation}
2\ln\mathcal{B}_{10} \approx \rho^2 + 2 \ln \left(\frac{\Delta V_1/V_1}{\Delta V_0/V_0}\right) \; ,
\end{equation}
where $\Delta V_\mathcal{M}$ is the characteristic spread of the likelihood around the maximum, and $V_\mathcal{M}$ is the total parameter space volume of the model.
The second term on the right-hand side is negative and imposes an Occam penalty, favoring models with fewer parameters.
While this expression is simple, in practice such a calculation requires detailed knowledge about the likelihood function for both the signal model and noise-only model. Current Bayesian PTA analyses use a nested model approach and a Savage-Dickey approximation to the Bayes factor, which does not furnish the noise-only likelihood function. Nonetheless, it is obvious from Eqns. (\ref{eq:weak_scaling}) and (\ref{eq:intermediate_scaling}) that one expects a monotonically increasing Bayes factor as the observation time for a PTA increases. 

One can relate the aforementioned scaling laws to an UL by using the complimentary error function,
\begin{equation}
A^2_{\rm UL}=\hat{A}^2_{\rm GWB}+\sqrt{2}\sigma_0 \;{\rm erfc}^{-1}\left[2\left(1-\epsilon\right)\right]
\end{equation}
where $\epsilon$ is the significance threshold for the limit, e.g. $0.95$ for a $95\%$ UL. The expectation value yields
\begin{align}
\left<A^2_{\rm UL}\right>=&A^2_{\rm GWB}+\sqrt{2}\sigma_0 {\rm erfc}^{-1}\left[2\left(1-\epsilon\right)\right]\\
=&A^2_{\rm GWB}\left(1+\frac{\sqrt{2}\; {\rm erfc}^{-1}\left[2\left(1-\epsilon\right)\right]}{\left<\rho\right>}\right).
\end{align}
Defining $\varepsilon\equiv\sqrt{2}\; {\rm erfc}^{-1}\left[2\left(1-\epsilon\right)\right]$, and given the time dependence of the SNR in Eqns (\ref{eq:weak_scaling}) and (\ref{eq:intermediate_scaling}), the ULs for the optimal statistic should evolve as:
\begin{subequations}
\begin{align}
\left<A^2_{\rm UL, weak}\right>&=A^2_{\rm GWB}+\frac{\varepsilon}{A^2_{\rm GWB}\delta_{\rm weak}}T^{-\frac{13}{3}}\; , \label{eq:weak_ul}\\
\left<A^2_{\rm UL, int}\right>&=A^2_{\rm GWB}+\frac{\varepsilon}{A^2_{\rm GWB}\delta_{\rm int}}T^{-\frac{1}{2}}\; , \label{eq:int_ul}
\end{align}
\end{subequations}
where $\delta_{\rm weak}$ and $\delta_{\rm int}$ are shorthand for the coefficients of time, $T$, in Eqns. (\ref{eq:weak_scaling}) and (\ref{eq:intermediate_scaling}), except for $A_{\rm GWB}$. 

Since these relationships are based on a frequentist statistic, it is prudent to compare them to the ULs obtained from a Bayesian analysis on simulated data sets, as it is well known that frequentist and Bayesian ULs can have different interpretations, see for instance \cite{Rover:2011zq}.

\subsection{Simulations}\label{subsec:simulations}
We simulated NANOGrav-like data sets following \cite{tss2017} using the Python wrapper for TEMPO2 \citep{Hobbs:2006_mnras}, and PTA simulation package, {\tt libstempo} \citep{vallisneri:2015}\footnote{\url{https://github.com/vallis/libstempo}}. 
We ran an UL analysis on each simulated data set with an injected GWB of known amplitude. The simulated data sets are based on the noise properties, and epochs of observation for the 11-year data set. An UL analysis was run for 200 different realizations of a GWB at $A_{\rm GWB}$ of $1\times10^{-16}$, $1\times10^{-15}$, and $3\times10^{-15}$. This analysis is identical to what we carry out in \S~\ref{sec:initial_slice_results}.

\begin{figure}[htp]\label{fig:sims_ng11yr_1e-15}

\includegraphics[clip,width=\columnwidth]{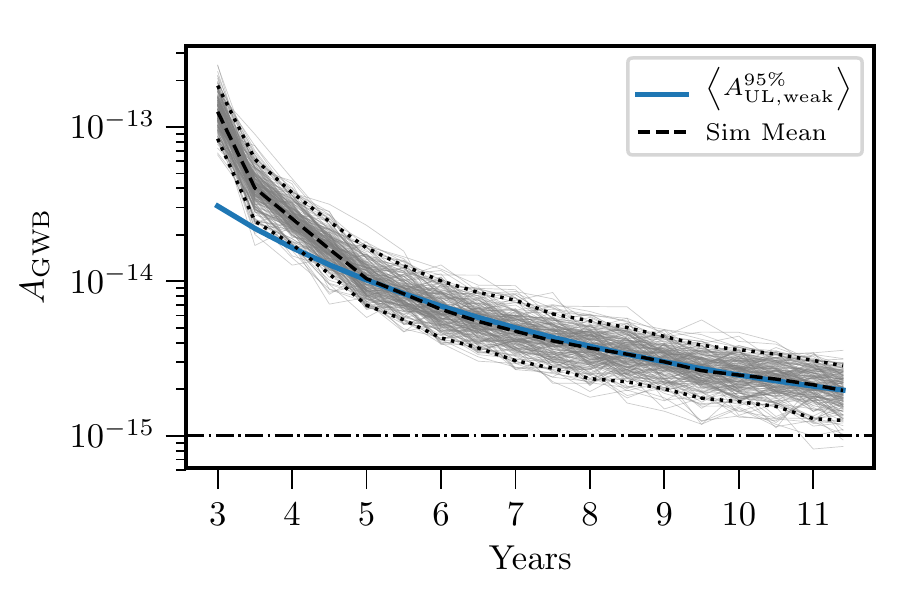}
 
\caption{The 95\% ULs over time for a set of NANOGrav-like datasets with an injection of $A_{\rm GWB}=1\times10^{-15}$. The mean UL and 90\% confidence intervals are shown for the 200 injections of a GWB. The blue curve is the predicted relationship between the UL, true value for the amplitude of the background and total time of observation. It has been fit to the mean values of the ULs in the range of $5-11.4$ years by varying $\delta_{\rm weak}$. }

\end{figure}

In \fig{fig:sims_ng11yr_1e-15} the results are summarized for the simulations with an injection of $A_{\rm GWB}=1\times10^{-15}$. The mean of the ULs and the $90\%$ confidence interval are shown, along with the level of the injection and a fit to the theoretical evolution for the UL in the weak regime, given by \eqn{eq:weak_ul}. The curve is fit to the mean values greater than $5$ years into the data set by varying the $\delta_{\rm weak}$ parameter.

\begin{figure}[htp]
\label{fig:sims_ng11yr_1e-16}
\subfloat{%
\includegraphics[clip,width=\columnwidth]{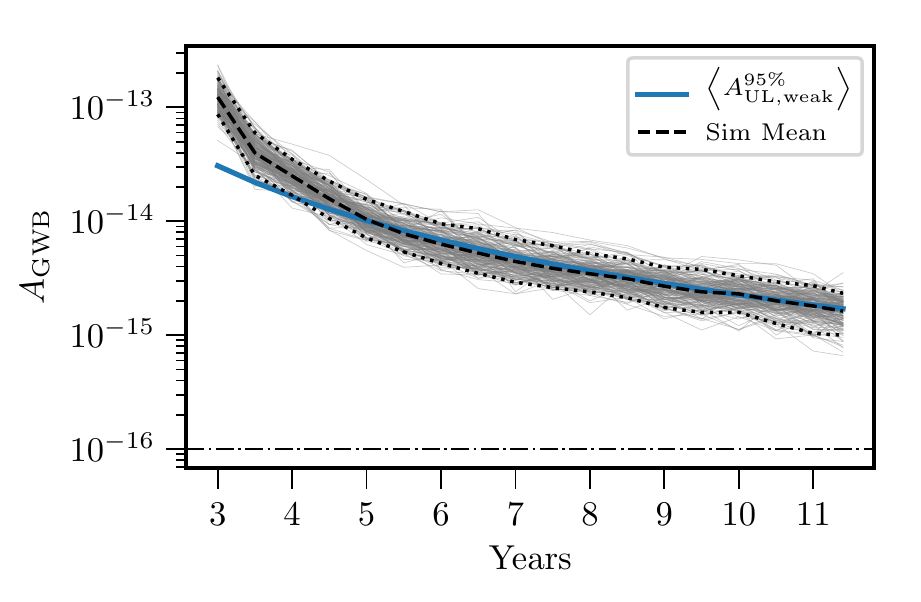}%
}

\vspace*{-0.2in}

\label{fig:sims_ng11yr_3e-15}
\subfloat{%
	
\includegraphics[clip,width=\columnwidth]{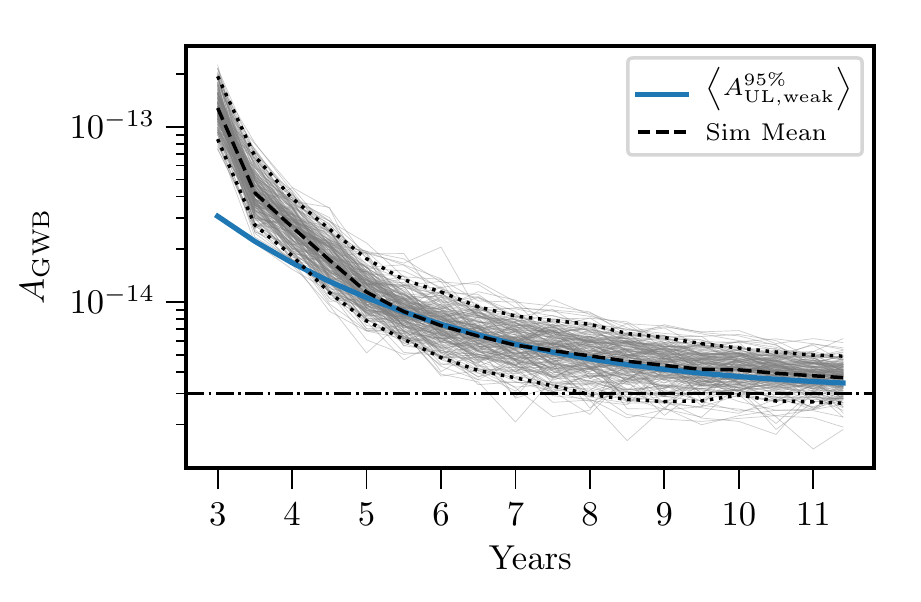}%
}

\caption{ The 95\% ULs over time for a set of NANOGrav-like datasets with injections of $A_{\rm GWB}=1\times10^{-16}$ and $A_{\rm GWB}=3\times10^{-15}$ are shown. The mean UL and 90\% confidence intervals are shown for the 200 injections of a GWB. The blue curve is the predicted relationship between the UL, true value for the amplitude of the background and total time of observation. Here the theoretical curve has {\bf not} been fit to the data. Rather we use the value for $\delta_{\rm weak}$ used in \fig{fig:sims_ng11yr_1e-15} and changed the value of $A_{\rm GWB}$ according to \eqn{eq:weak_ul}. }
\end{figure}

Such a close fit from $5-11$ years is {\it extraordinary} given that many of the other parameters in this relationship are changing with time, e.g., the average cadence and average TOA error. In part, it is these large changes in parameters at the beginning of the data set that are responsible for the poor fit to the theoretical prediction at early times. 
%%%%%%%% NG11yr Sims A=1e-15 %%%%%%%%%%
This fit for $\delta_{\rm weak}$ is then used in \fig{fig:sims_ng11yr_1e-16} to compare the theoretical evolution of \eqn{eq:weak_ul} to the evolution of the UL with larger and smaller injections of the GWB. With the same value for $\delta_{\rm weak}$, only changing the injection strength, $A_{\rm GWB}$, accordingly, the Bayesian analyses of these simulations follow the theoretical predictions at late times. 
%%%%%% NG11yr Sims A=1e-16, 3e-15 %%%%%%%%

Armed with a general understanding of the expected evolution of GWB analysis statistics, we move on to the sliced analysis of the NANOGrav 11-year data set.

%%%%%%%%%%%%%STANDARD ANALYSIS RESULTS%%%%%%%%%%
\section{Standard Analysis Results}
\label{sec:initial_slice_results}
Here we report the results of both our standard detection and UL analyses of the NANOGrav 11-year data set for a GWB across the sliced data set. 

The expectations laid out in \S~\ref{sec:motivation} are that in this type of analysis, as more data are added (i.e., we have more information about the system we are studying) the posteriors for $A_{\rm GWB}$ should get narrower. 
In broad strokes this means that the Savage-Dickey Bayes factor approximation would start near to one and begin to increase as more data are added. The SNR should also increase according to the evolution described in \citet{2013CQGra..30v4015S}, while the UL should steadily decrease with more data until the data become sufficiently informative as to run up against the actual signal amplitude, as shown in \S~\ref{sec:motivation}.
{\it In both of our Bayesian analyses, the evolution of these statistics does not conform to these reasonable predictions.}

In the case of a varied-spectral index analysis we expect similar behavior, but as a detection becomes imminent, the significance of a steep-spectral index common process will increase. {\it In the case of the varied-spectral index analysis, rather than seeing a steep spectral index near $\gamma=13/3$, a shallow $\gamma\approx 0 $ process appears early in the observation period.}

\subsection{Fixed Spectral Index Analyses}
\begin{figure}[tp!]
\begin{flushleft}
\label{fig:std_ul}
\subfloat{%[Fig6a.pdf][The 95\% upper limits for cumulative years of data.]
  \includegraphics[clip,width=0.95\columnwidth]{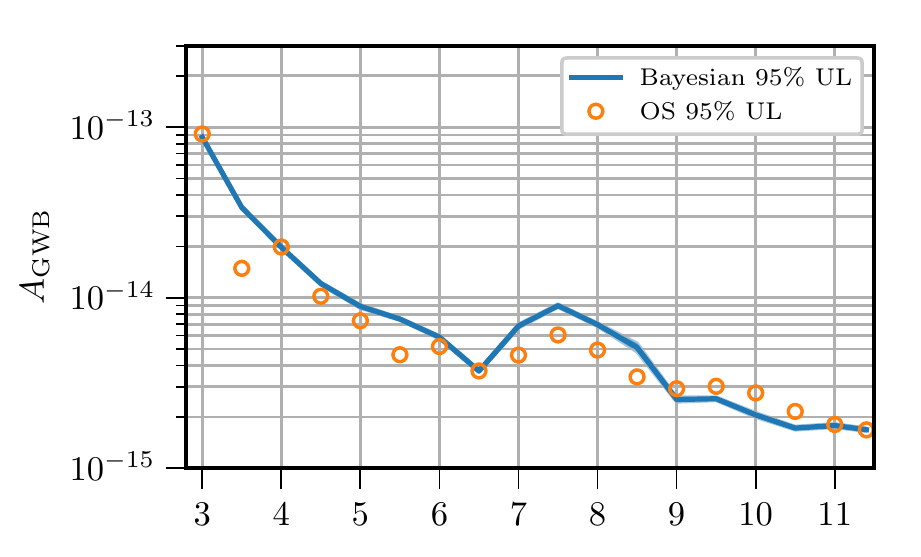}%
}
\end{flushleft}
%\centering

\label{fig:std_bf}
\subfloat{%[Fig6b.pdf][Bayes Factors]
  \includegraphics[clip,width=\columnwidth]{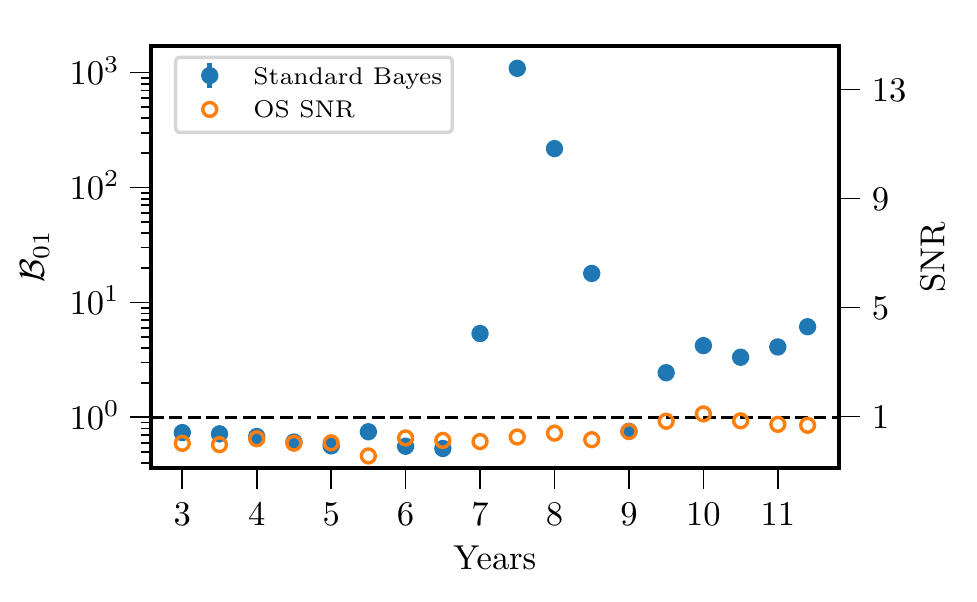}%
}

\caption{Results of a standard PTA gravitational-wave data analysis. In both plots the blue solid line/dots show the Bayesian results at each slice, while the open orange circles show the frequentist results. Note the large rise in the UL and the Bayes factor for the Bayesian analysis starting in the $6.5$-year slice and peaking in the $7.5$-year slice. The UL calculated from the noise-marginalized optimal statistic has a similar trend as the Bayesian UL, but with lower values. The SNR calculated from the optimal statistic shows no sign of a detection in the same era.}

\end{figure}

The Bayes factor in the bottom panel of \fig{fig:std_bf}, remains between $0.5$ and $1.5$ until $7.0$ years into the data set, then increases dramatically for a few slices, before decreasing again.
As can be seen in the top panel of \fig{fig:std_ul} the UL decreases monotonically until six and a half years into the data set (slice ending about MJD 55590 (2011.08)) and sharply increases over the next year until the 7.5-year slice (slice ending about MJD 55956 (2012.08)), before beginning to decrease again. We refer to this period of time as the \kink\ for brevity. 

The frequentist statistics are mixed. While the UL calculated from the optimal statistic noise-marginalized posteriors are a bit lower during the \kink, the same trend (particularly the increase) in the UL can be seen centered around the 7.5-year slice. The SNRs are, however, drastically different from the Bayes factor trends, and do not show the dramatic increase in this era. Since these calculations involve spatial correlations they are often used as the quickest estimate of detection capabilities of a given data set. The fact that they are so different in this era from the Bayesian results is troubling and is the impetus for most of the remainder of this paper.

%%%%%% Standard Analysis UL/BF %%%%%%%%

A full sliced analysis using Hellings-Downs spatial correlations was undertaken on the data set. These analyses hint at an even stronger detection of a GWB. The posteriors for this analysis indicate a GWB detection too strong to be estimated using the Savage-Dickey Bayes factor approximation. Since these analyses take up to ten times longer to run we restricted our follow-up analyses to searching for an uncorrelated common red process among all pulsars.  

\subsection{Varied Spectral Index}
\begin{figure}[htp]\label{fig:vary_gamma1}

\includegraphics[clip,width=\columnwidth]{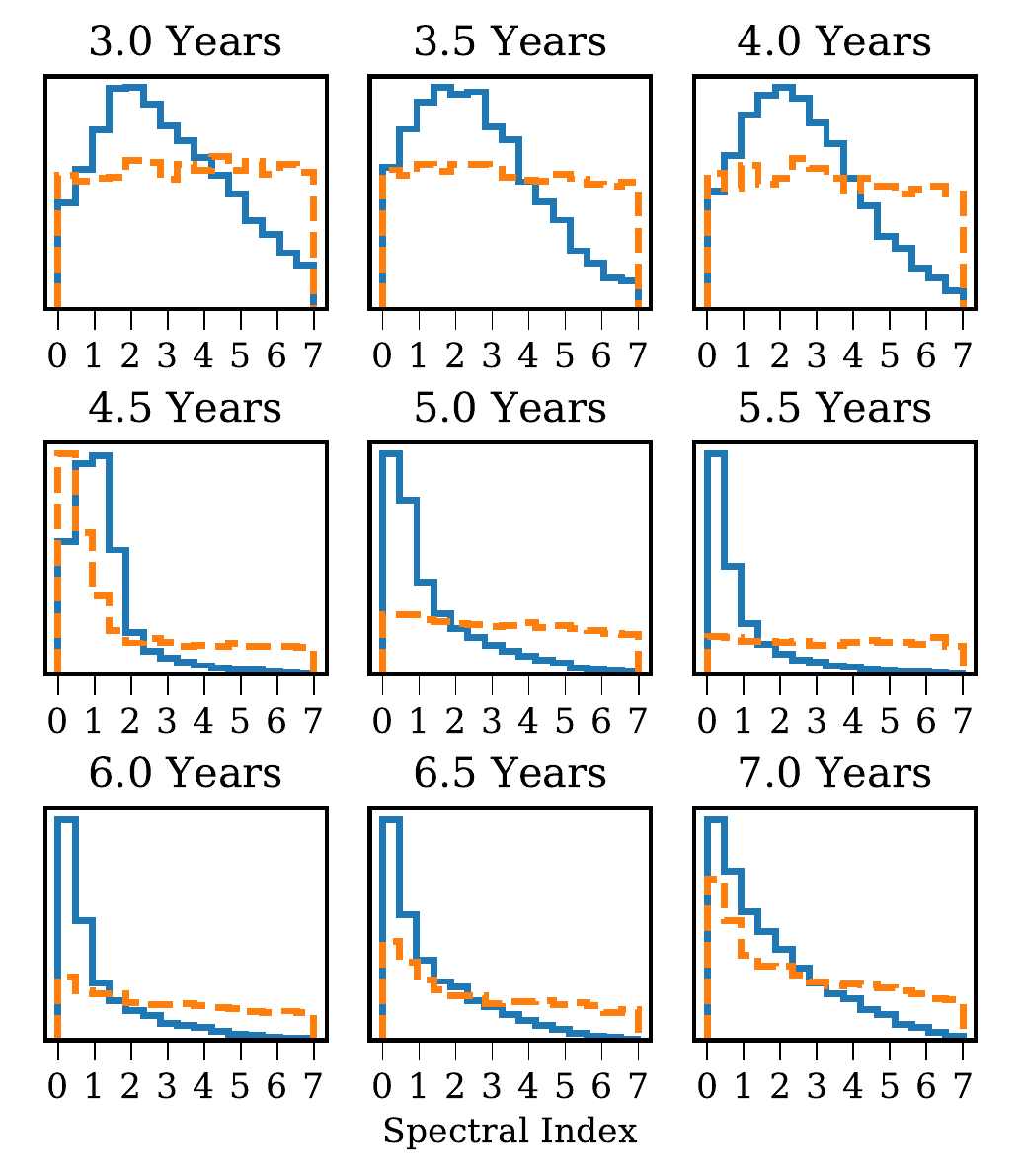}
 
\caption{Comparison of posteriors for the spectral index of a power law red noise process across the first six slices. The solid blue line is the posterior on the spectral index for a common process between all pulsars. The dashed orange line shows the posterior for the spectral index in the single pulsar red noise model for PSR J1713+0747. Note the appearance of a strong low-spectral index in both posteriors for the 4.5-year slice.}

\end{figure}

One other analysis that needs to be summarized is that with a varying spectral index for the common red process model for the GWB.
This analysis shows a drastic change in the posteriors of the parameters, but this change is {\it not} contemporaneous with the \kink\ in the analyses where the spectral index is set to $\gamma=13/3$. 
Here we see that the spectral index describing the common red noise between the pulsars changes dramatically in the $4.5$ year slice ending at MJD 54860 (2009.08) and butts up against $\gamma=0$, \fig{fig:vary_gamma1}.

The feature slowly dissipates until roughly the 7.0-year slice. 
This change in the spectral index {\it is} contemporaneous with the ``first'' chromatic timing event in PSR J1713+0747 \citep{dfg+13,Lam:2017duo}. Investigating further we see that a similar feature exists in the single-pulsar red noise analysis for this pulsar, the orange-dashed curve in \fig{fig:vary_gamma1}. We will return to this feature and investigate this correlation with the ``first'' PSR J1713+0747 ISM event in \S~\ref{sec:lowspec}.

%%%%%%%% Compare Varying Gamma across early Slices %%%%%%%%%%

Additionally, the recovered value of $\gamma$ coincident with the \kink\ does not significantly vary. Rather then moving towards larger values, indicative of the recovery of a ``steeper'' process, the spectral index recovery remains broad and stagnant. 
While this is in no way dismissive of a GW-related event occurring during this time, it is further evidence of the anomalous behavior of our Bayesian analyses throughout the \kink.

%%%%%%%%%%%%%%INVESTIGATING THE ANOMALOUS RESULTS%%%%%%%
\section{Investigating the Anomalous Signal}
\label{sec:results}
In this section we investigate the anomalous GWB signal in our sliced analysis, which includes the \kink\ in the UL analysis and the large spike in Bayes factor contemporaneously. We run through a number of the diagnostic analyses that were completed and summarize our mitigation strategy.

\subsection{Slice-Specific Simulations} \label{subsec:slice_simulations}

While the evolution of our UL and Bayes factors seems to be unexpected, these type of statistics are expected to evolve stochastically depending on how the particular noise and Gaussian process realizations interact with the data. This is evident from the simulations run in \S~\ref{sec:motivation}. Therefore, it is no surprise that the UL or Bayes factor is non-monotonic in moving from any given slice of the data to the next. The parameter space for a PTA is large and the data sets for the individual pulsars may interact in complex ways with various lengths of observation, red noise frequencies used and white noise parameter characterization. However, such a large and continuous rise in the Bayes factors and ULs seems to be worth further investigation.

The \kink\ in the UL time series of \fig{fig:std_ul} lies far outside of the $90\%$ confidence intervals of the simulations in \S~\ref{sec:motivation}. Noting that noise parameter characterization can change substantially over time, this large deviation inspired a {\it new} set of simulations, in order to better characterize the degree to which the \kink\ is just a statistical fluctuation. This set of simulations uses the same techniques of \S~\ref{subsec:simulations} with one major difference. Rather than use the noise parameters from the full 11-year analysis to simulate the pulsar data sets, the noise parameters retrieved from a single-pulsar analysis {\it at each slice} were used to build a simulation of that slice. This gave us an injection that was more ``true'' to the knowledge at the time of a particular slice. In practice this involves making a whole new data set for each slice. This analysis was done on a set of data sets with an injection of $A_{\rm GWB}=10^{-15}$.

\begin{figure}[htp]\label{fig:sims}

\includegraphics[clip,width=\columnwidth]{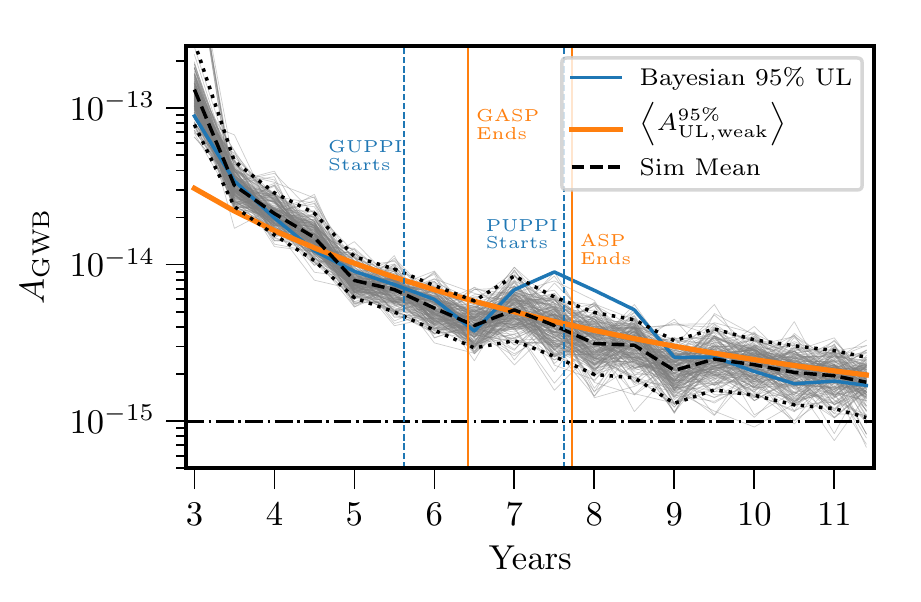}
 
\caption{UL analyses of simulated NANOGrav 11-year-like data sets with slice-dependent noise characteristic. Data was simulated for each slice based on the noise parameters recovered from the sliced analyses on the real data set. This is in contrast to the simulations done in \sect{sec:motivation}, where the noise parameters from the full data set were used to simulate a full data set which was then sliced for the UL analyses. One can see that the difference in noise parameters makes the \kink\ of the blue trace somewhat less significant. The orange line shows the expected evolution, and is identical to the blue line in \fig{fig:sims_ng11yr_1e-15}. The vertical lines show when the new GUPPI/PUPPI backends came into use and when the old GASP/ASP backends were phased out.}

\end{figure}

The mean UL and 90\% confidence interval are plotted in \fig{fig:sims}.
%%%%%%%% Sliced-Noise Simulations %%%%%%%%%%
Besides the mostly monotonic trend in the average UL, there is a characteristic plateau that starts just around the increase in the real data.  This plateau appears near the changeover in our pulsar backend at the GBT from GASP to GUPPI. GUPPI and PUPPI allow for much wider band observations of radio pulses, which allows us to do more accurate mitigation of dispersion measure fluctuations. While no causal relation has been found between this changeover and the \kink\ we will present a number of results in \S~\ref{sec:misc} that summarize further investigations.

Even with this plateau in the same era as the \kink\ the values of the UL in the $7.5$-year slice are outside of the $90\%$ confidence interval of the simulations done. The fact that the \kink\ is lessened in significance when simulating with noise estimated from each slice suggests either significant covariance between noise and GW signal at these epochs, or non-stationary noise features. The results from these simulations, along with the large Bayes factors, prompted us to examine which pulsars seemed to be most responsible for the signal. 

\subsection{Dropout Analysis}\label{sec:dropout}

It is important to point out that a number of different types of GW signal search are done by the NANOGrav collaboration on each new data set. In addition to searching for various types of stochastic backgrounds, a search for single sources of GWs from SMBBHs \citep{Aggarwal:2018mgp} and a search for GW memory \citep{Aggarwal:2019b} from binary coalescences\footnote{A rudimentary version of a generic burst search, based on the signal model and search algorithm of \citep{Ellis:2016mtg}, was also done on this data set in the course of these investigations with no significant evidence for GWs.} were performed. The most recent studies involving these searches cover both the \citetalias{abb+15} and \citetalias{abb+17} data sets. In both cases there was mild evidence for GW signals in \citetalias{abb+15}, which decreases in the analyses of \citetalias{abb+17}. In the case of the single-source search of \citet{Aggarwal:2018mgp} PSR J0613$-$0200 was found to be responsible for the spurious signal and both PSRs J0030+0451 and J1909$-$3744 were responsible for the burst-with-memory anomaly in \citetalias{abb+15}. As mentioned in those manuscripts, both noise in the individual pulsars and the low-sensitivity sky positions of PSRs J0030+0451 and J0613$-$0200 were to blame. We will see that these pulsars again appear as culprit pulsars in this analysis. These pulsars were tagged as being the source of the spurious signals using a Bayesian dropout analysis. In none of the cases above was a robust detection of GWs found.

Following \citet{Aggarwal:2018mgp} a dropout analysis was undertaken on the sliced data set. The dropout analysis is a new technique for isolating spurious noise sources from particular pulsars in a PTA gravitational wave analysis \citep{Aggarwal:2018mgp}.
In a dropout analysis the signal being analyzed is coded with a so-called {\it dropout parameter}. 
These parameters multiply the signal amplitude and are sampled in the MCMC analysis. 
They are binary in the sense that depending on the sample the dropout parameter is either one or zero; turning the signal on/off. 
This allows one to use the Bayesian analysis to determine the signal model in terms of which pulsars prefer the signal to be turned on in the analysis. 
See \citet{Vigeland:2019} for more details. 

\begin{figure}[htp]
\label{fig:rm_culprits_ul}
\subfloat{%[Fig6a.pdf]
\includegraphics[clip,width=\columnwidth]{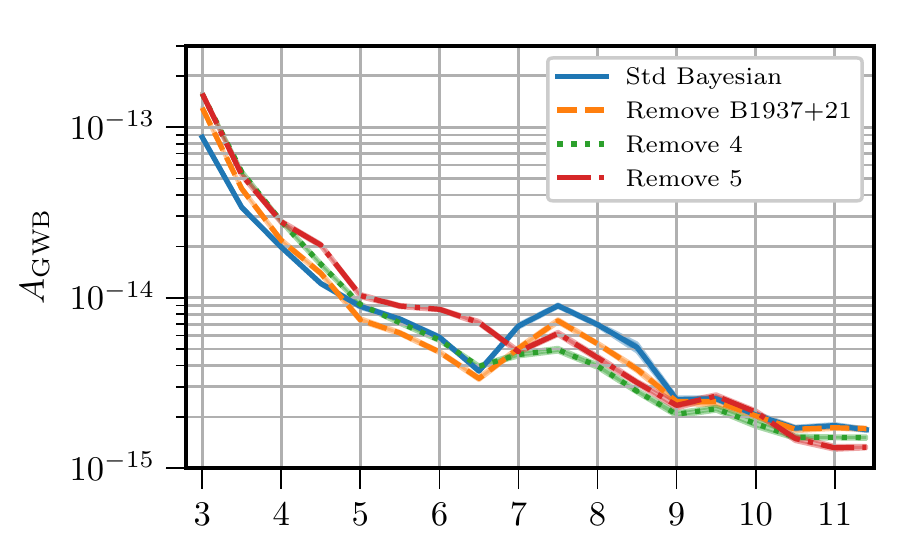}%
}

%\centering
\begin{flushright}
\label{fig:rm_culprits_bf}
\subfloat{%[Fig6b.pdf]
	
\includegraphics[clip,width=0.95\columnwidth]{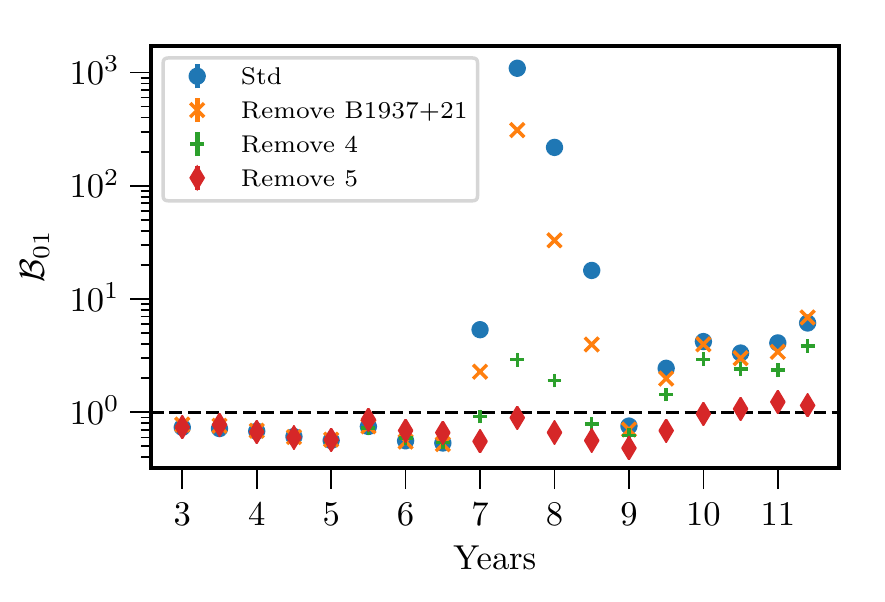}%
}
\end{flushright}
\caption{ULs and Bayes factors are shown for sliced analyses of the NANOGrav 11-year data set where some combination of culprit pulsars were removed.  The solid-blue line and blue dots are from the standard Bayesian analysis and are the same as in \fig{fig:std_ul}. The orange-dashed line and orange $\times$'s are from an analysis where B1937+21 has been removed. The green-dotted line and green +'s are from an analysis where B1855+09, B1937+21, J0030+0451, J0613$-$0200 have been removed. The red-dot-dashed line and red diamonds are from an analysis where B1855+09, B1937+21, J0030+0451, J0613$-$0200 and J1909$-$3744 have been removed.}

\end{figure}

Here we have used the GWB dropout analysis in {\tt enterprise\char`\_extensions} to look at which pulsars favor a common red signal across the slice analysis. 
The only pulsar with an odds ratio significantly higher than one in the $7.5$-year slice (i.e., more samples favor the presence of a stochastic background) is PSR B1937+21. 
It has long been known that this pulsar has a great deal of red noise \citep{ktr94}, so much that it was not included in the analysis by \cite{abb+16}. 
As can be seen in \fig{fig:rm_culprits_bf}, removing this pulsar from the analysis decreases the Bayes factor and the UL during this era. 
Therefore {\it some} of the spurious signal in the \kink\ era is due to this pulsar. However, while the UL and Bayes factors decrease across this set of time slices, the main features of the \kink\ are still present in the statistics.
%%%%%% Remove Culprits from UL/BF analyses %%%%%%%%

\subsection{Single-Pulsar GWB Statistics}\label{sec:single_psr_gw}

In search of other pulsars that might be responsible for this spurious signal, an exhaustive analysis of the GWB statistics was done for each individual pulsar. This analysis is often carried out to characterize the robustness of PTA gravitational-wave statistics and has been used in the past to rank the sensitivity of pulsars to the GWB. These individual pulsar ULs are in turn used to do cumulative analyses where pulsars are added until the UL asymptotes to a stable and more robust value. 

\begin{figure}[htp]\label{fig:single_psr_bf}

\includegraphics[clip,width=\columnwidth]{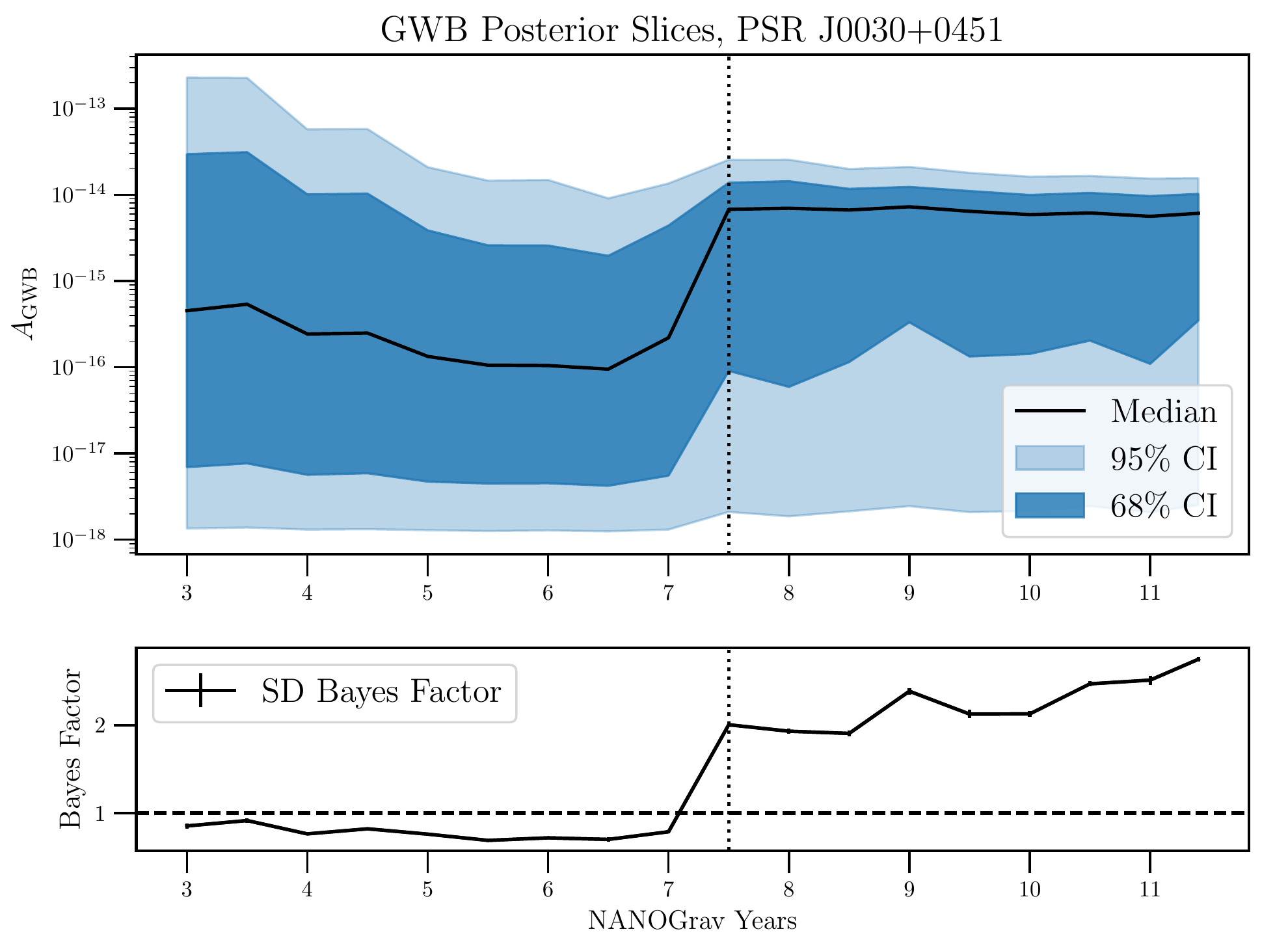}
 
\caption{Evolution of the GWB statistics in PSR J0030+0451. The top panel shows the median, 68\% and 95\% confidence intervals for the posterior on $A_{\rm gwb}$ at each slice. The bottom panel shows the Savage-Dickey Bayes factor calculated at each slice. Note, in both cases, the jump that occurs at the 7.5-year slice.}

\end{figure}

In an attempt to track down pulsars that could possibly be culprits in causing the \kink\ we ran individual Bayes factor and UL analyses on all 34 pulsars in the GW analysis, over the slices in question. 
This was not done initially because it is computationally intensive, requiring upwards of a thousand individual analyses. 
Plots of all of the pulsars' statistical evolution can be found online\footnote{\url{http://data.nanograv.org/static/data/ng11yr_slice_noise_plots.tar}}, but \fig{fig:single_psr_bf} summarizes those of one interesting candidate to be a source of the anomalous statistics, PSR J0030+0451. Four culprit pulsars were identified in these analyses: PSRs B1855+09, J0030+0451, J0613$-$0200, and J1909$-$3744.
%%%%%%%% Single Pulsar Bayes Factor %%%%%%%%%%
These pulsars all show either a similar feature to the full PTA analysis or a sharp rise in the Bayes factor in the 6.5 year to 8.5 year time span. 
This same feature does not appear in the UL and Bayes factor time series for PSR B1937+21 but we group it with these pulsars since it affects the GWB statistics in this era, as determined by the dropout analysis. 

After finding these candidates, the most straight forward test of their responsibility for the \kink\ was to remove each of them and run the analysis on the remaining pulsars in the PTA. 

Analyses were done with each of the pulsars removed individually and in all subsets of the culprit pulsars. We only discuss the results of removing either all of the pulsars, or four of the pulsars while keeping PSR J1909$-$3744 in the analysis, since these are the most interesting cases. 

\fig{fig:rm_culprits_ul} shows these cases. 
When four of the culprits (not including PSR J1909$-$3744) were removed from the analysis the \kink\ strongly decreases in the time series and the evolution of the UL falls well within the $90\%$ confidence interval of the simulations in \fig{fig:sims}. 
The Bayes factor time series still shows a remnant feature during this period but when PSR J1909$-$3744 is removed the Bayes factor time series decreases even further. The removal of this pulsar drastically reduces our sensitivity to the GWB across the timespan of interest. This is not surprising since PSR J1909$-$3744 is one of our most precisely timed pulsars but in the next section it will be important to include this pulsar in our mitigation strategy.

\subsection{Free Spectral Red Noise Models}\label{sec:fs_rn_model}
While it is important to identify which pulsars have such strong, spurious GW detections, we are primarily interested in finding a way of mitigating the noise in these pulsars so they can be included in the full PTA analysis. 
In \S~\ref{sec:single_psr_gw} we demonstrated that removing a small subset of the pulsars from the GW analysis removed the \kink. 
While it is reassuring that we have isolated the spurious detection of a GWB to a handful of pulsars, we would like to devise a mitigation strategy for this type of noise and understand the root cause more thoroughly, especially since one of our most sensitive pulsars is included in the list. 

As discussed in \S~\ref{sec:data_analysis}, a free-spectral model is another tool in the standard PTA data analysis toolbox that we can use to model the red noise in these individual pulsars. 
The free spectral model uses a free parameter for the amplitude of the Fourier basis red noise at each frequency modeled, which provides a much larger parameter space for the noise by not restricting the noise in a given pulsar to follow any type of functional dependence in frequency. 
Hence, such a model can model the noise in the lowest frequency bins of an individual pulsar {\it and} any unmodeled higher frequency noise independently, which the power law model is unable to do.

In order for the free spectral model to cover the same number of frequencies as the power law model upwards of 30 parameters need to be added per pulsar. 
Historically, this is the primary reason for not including free spectral models in PTA noise analyses, as the increased parameter space becomes computationally infeasible to search over 
Additionally, forthcoming work \citep{Simon:2019} shows that these models do not compete well in a Bayesian model selection framework because the large Occam penalty of the additional parameters cancels out the ability of these many parameters to describe the noise accurately. 

\begin{figure}[htp]
\label{fig:fs_ul}
\subfloat{
  \includegraphics[clip,width=\columnwidth]{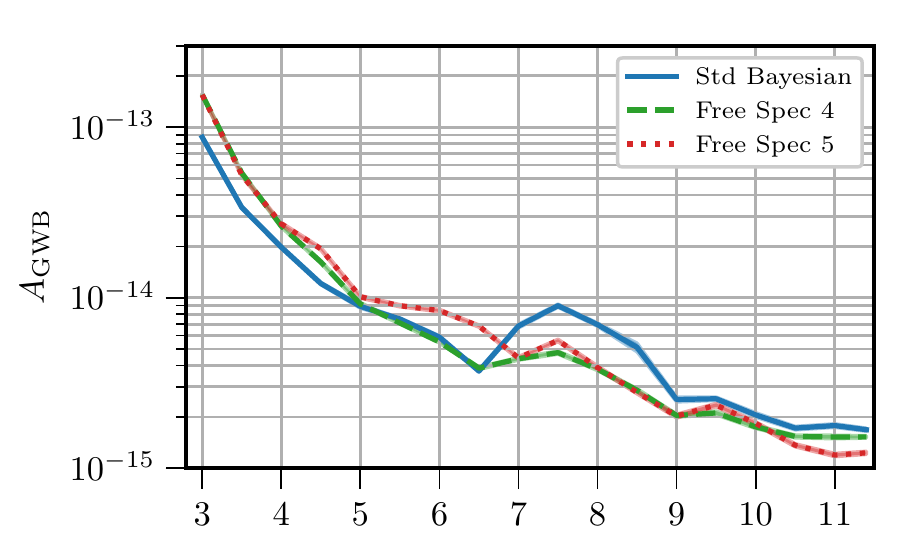}%
}

%\centering
\begin{flushright}
\label{fig:fs_bf}
\subfloat{
	
  \includegraphics[clip,width=0.95\columnwidth]{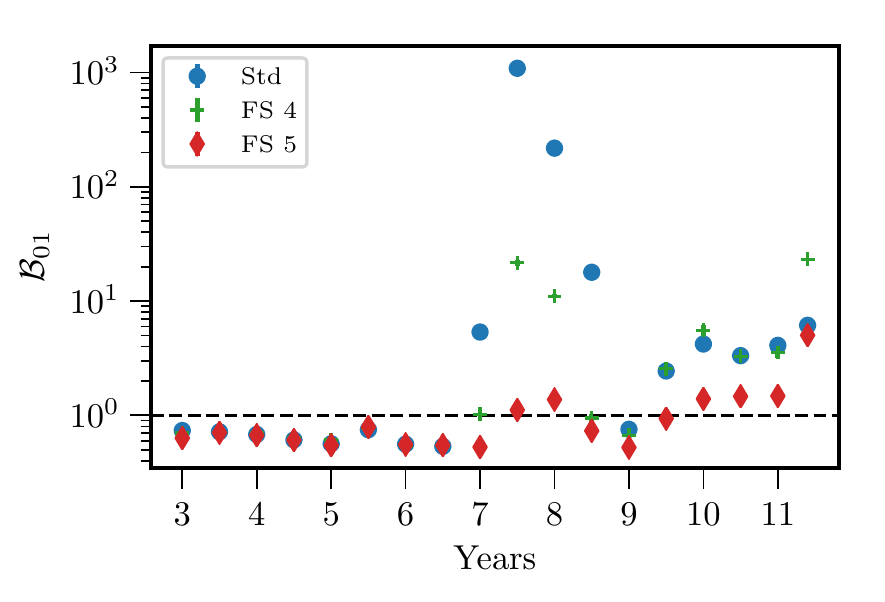}%
}
\end{flushright}
\caption{ULs and Bayes factors are shown for sliced analyses of the NANOGrav 11-year data set where some combination of culprit pulsars were analyzed with a free-spectral-noise model. The solid-blue line and blue dots are from the standard Bayesian analysis and are the same as in \fig{fig:std_ul}. The green-dotted line and green $+$'s are from an analysis where PSRs B1855+09, B1937+21, J0030+0451 and  J0613-0200 have free-spectral-noise models. The red-dot-dashed line and red diamonds are from an analysis where PSRs B1855+09, B1937+21, J0030+0451, J0613$-$0200 and J1909$-$3744 have free-spectral-noise models. Note the differences in the 11.4-year slice for these analyses, as compared to \fig{fig:rm_culprits_ul}.}

\end{figure}

With those caveats in mind, we undertook another set of PTA gravitational wave analyses using free spectral models for the culprit pulsars. The results are summarized in \fig{fig:fs_ul}. 
These results demonstrate that the free spectral model is effective at mitigating the spurious noise features in both the UL and Bayes factor analyses of the NANOGrav 11-year data set. 
It is salient to compare the longest slices, i.e., $>10$ years, where the Bayes factor is slowly increasing. 
With all of the culprit pulsars removed in \fig{fig:rm_culprits_bf} the Bayes factor still remained at or below one in these late slices, showing no early signs of any type of signal. However, even though the use of a free spectral noise model for the five culprit pulsars mitigates the spurious features identified in this paper, it begins to favor the signal model in the long run, focusing on the points $>10$ years in \fig{fig:fs_bf}. If this growth in the Bayes factor were the beginning indications of a real signal, its growth would be indicative of the amplitude of the underlying GWB. A separate, ongoing investigation is addressing this question by injecting GWB signals into the $11$-year data set and will be published separately.
%%%%%% Free Spectral UL/BF analyses %%%%%%%%

\begin{figure}[htp]\label{fig:asd_7p5yr_J0030+0451}

\includegraphics[clip,width=\columnwidth]{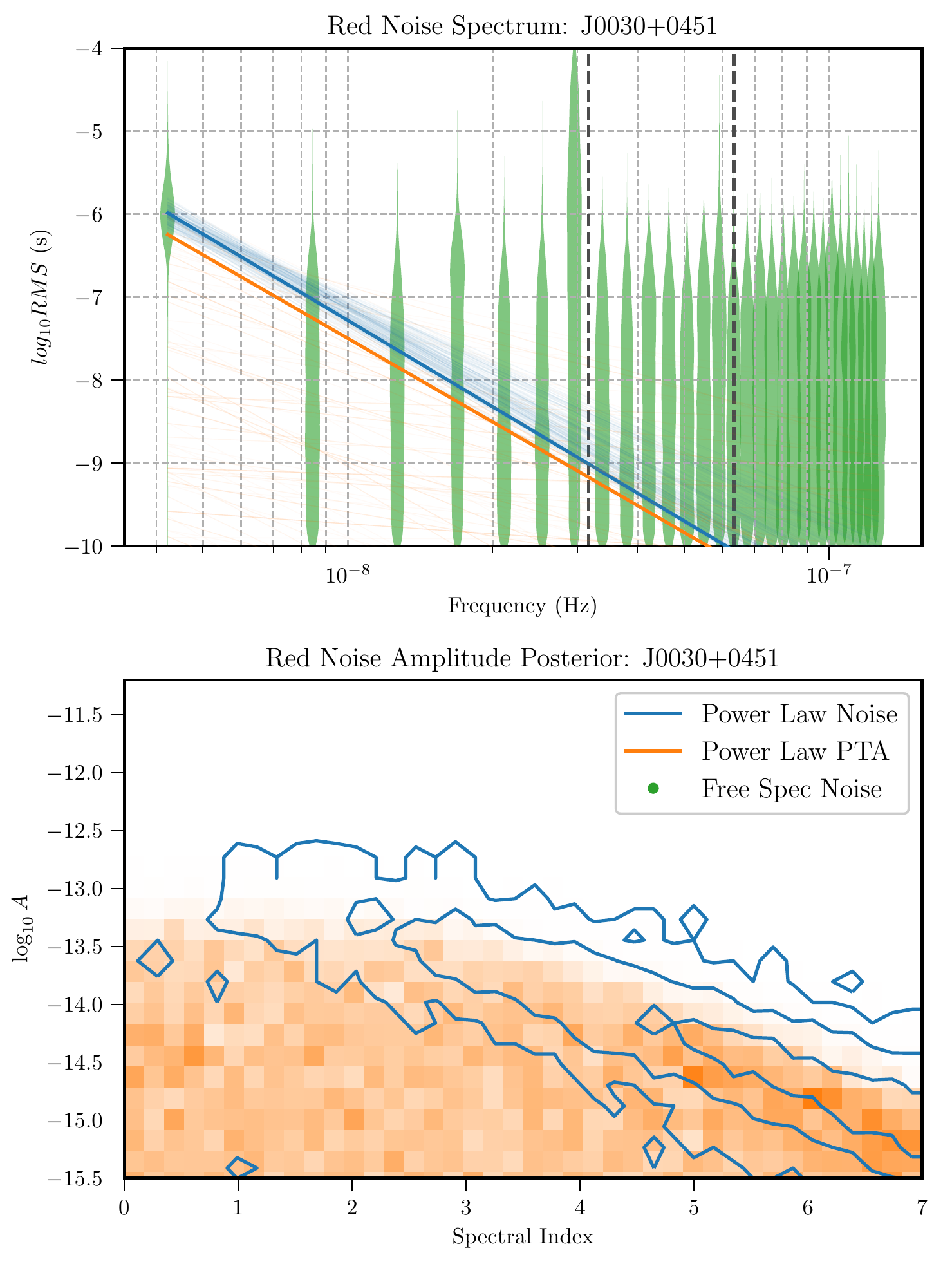}
 
\caption{Noise model posteriors for PSR J0030+0451 at the 7.5-year slice. The top panel shows realizations of the power law red noise as straight lines in the log-log plot. The heavy traces are the power law for the maximum likelihood values, while the fainter traces show a representative sample of power law realizations from the Bayesian analysis. The legend in the bottom panel is also accurate for the top panel. Power law red noise realizations from the individual noise analyses (blue) and full PTA analysis (orange) are shown. The (green) violin plots show the posteriors for the free-spectral-noise model at each frequency. One can judge the significance of a detection by how separated the violin plot is from the lowest amplitudes. The crowding at higher frequencies stems from the linearly spaced frequencies on a log-scale. The vertical-dashed lines show frequencies at $1/{\rm yr}$ and $2/{\rm yr}$. 
The bottom panel shows the 2-dimensional posteriors for the power law-noise models, in $\gamma$ and $\log_{10}A_{\rm GWB}$. The blue contours show the posterior from the individual noise run, while the orange heat map shows the posterior for the full PTA run for the same pulsar, PSR J0030+0451. Note that while the individual pulsar noise run shows closed contours for the power law model, the full PTA has a very diffuse, non-significant posterior. It is suspected that this RN power has moved into the common red noise process, which shows a strong detection in the 7.5-year slice.}

\end{figure}

Red noise amplitude spectral densities for the $7.5$-year slice are shown in \fig{fig:asd_7p5yr_J0030+0451} for PSR J0030+0451.
The free-spectral parameter posteriors are compared to a sample of the power law posterior amplitude spectral densities. The bold straight lines are the power law spectrum for the maximum likelihood values for these power law parameters.
The second panel shows the two-dimensional posteriors for the power law parameters, in both the single pulsar noise run and the full PTA analysis. 
In the first panel note that the only significant free-spectral parameter (i.e., sufficiently separated from the minimum amplitude) is the one for the lowest frequency.
This points to a possible cause for the anomalous signal we are seeing from a few pulsars.
The lowest frequency will model all power within a $\delta f$ defined by the inverse of the time span but it is limited by the second-lowest frequency where there is no substantial evidence for power.
One conjecture, partly substantiated by comparing the different two-dimensional power law posteriors, is that the power law is able to find this power at low frequencies, but since the signal is only significant at one frequency, this power is allowed to transfer between the pulsar red noise model and the GWB common red process. 

\begin{figure}[htp]\label{fig:asd_11p4yr_J0030+0451}

\includegraphics[clip,width=\columnwidth]{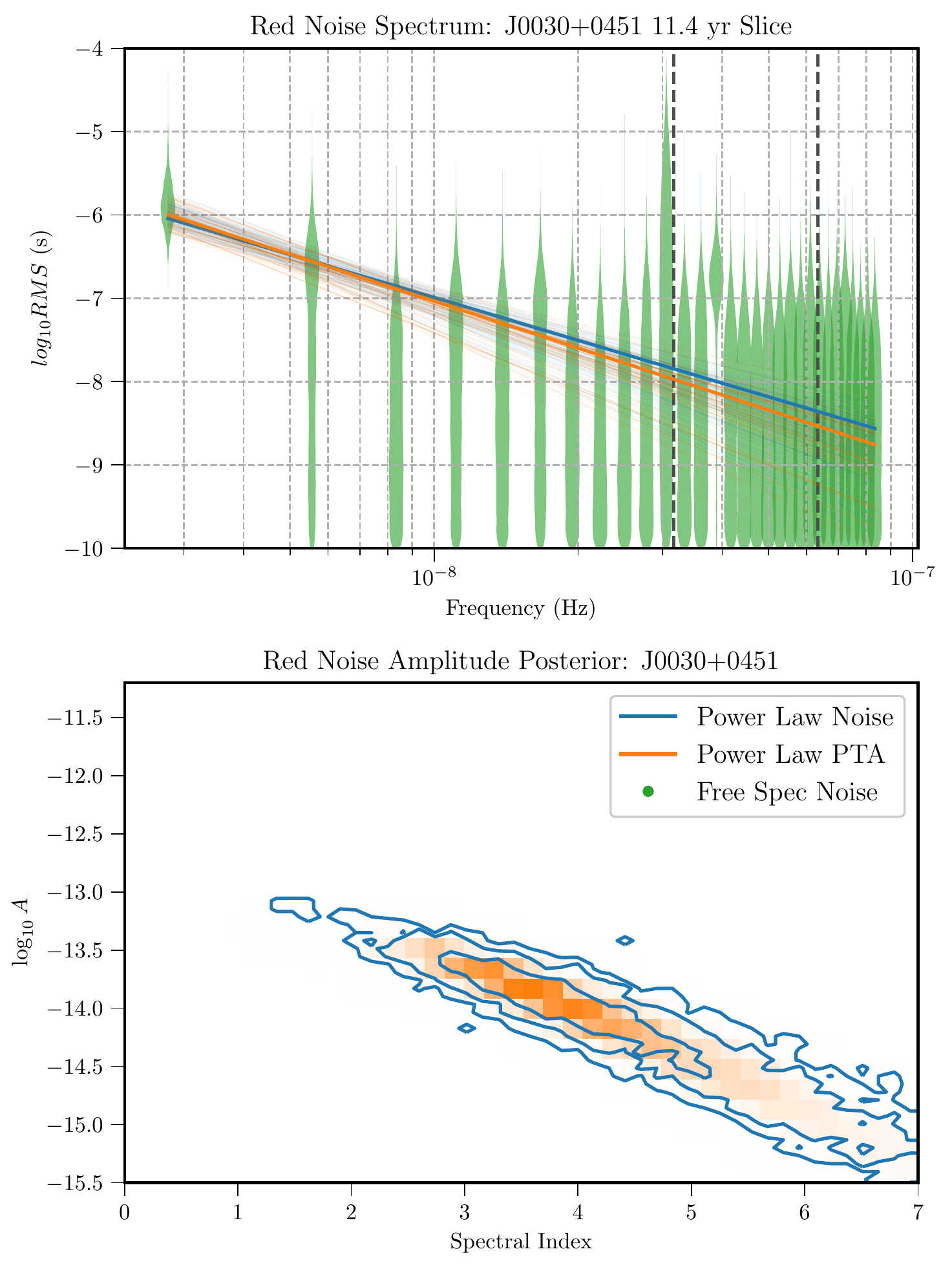}
 
\caption{Noise model posteriors for PSR J0030+0451 for the full data set. The various elements are the same as in \fig{fig:asd_7p5yr_J0030+0451}, but the results are reveal how the \kink\ is in part mitigated with time. Looking at the upper panel, in addition to the lowest frequency, the second-lowest frequency posterior for the free-spectral-noise model is also above the WN floor. The power law model is not as hindered by the WN floor, and as can be seen in the lower panel, the power law-noise model effectively holds much of the red noise power in this pulsar, rather than allowing it to all move into the common red noise process.}
\end{figure}

Compare these results to the same data products from the full NANOGrav 11-year data set in \fig{fig:asd_11p4yr_J0030+0451}. The second-lowest frequency free-spectral parameter is more significant, and the power law model is much more consistent between the single pulsar noise analysis and the full PTA analysis.
%%%%%%%% Compare at 7.5 Year Slice %%%%%%%%%%

%%%%%%%% Compare at 11.4 Year Slice %%%%%%%%%%

%%%%%%%%%%%%TOY MODEL %%%%%%%%%%%%%%%%
\subsection{Anomalous Signal Toy Model}
\begin{figure}[htp]\label{fig:j0030_sims_bf}

\includegraphics[clip,width=\columnwidth]{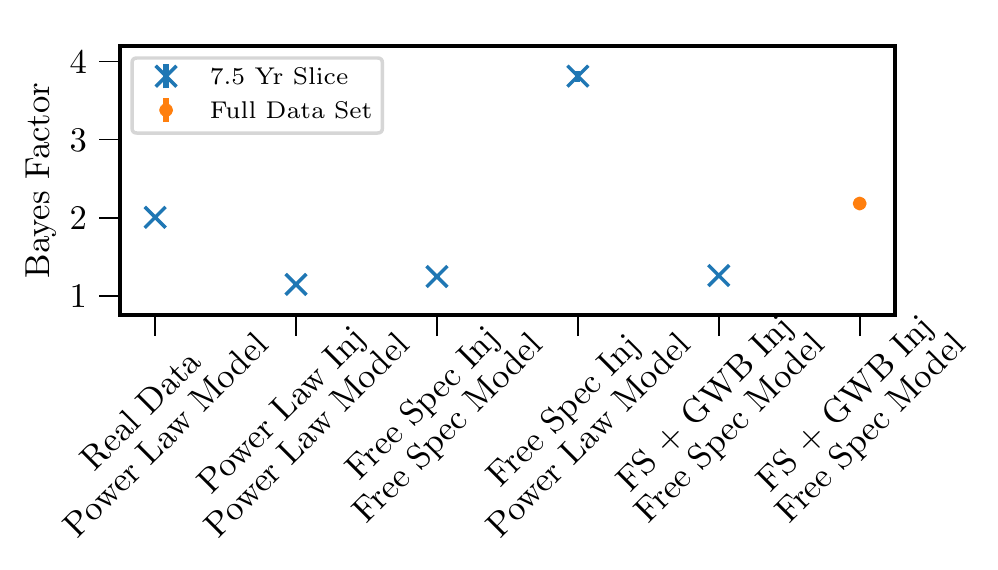}
 
\caption{Bayes factor for the GWB amplitude from various noise injections and analyses. Error bars are included, but are smaller than the markers in most cases. The labels show what type of injection and noise model were used. Note that using a free-spectral model injection while using a power law model in the analysis results in a higher significance detection of a GWB.}

\end{figure}

To try and further understand the spurious signal seen in the $7.5$-year slice we ran a number of simulations using PSR J0030+0451-like data sets in an attempt to duplicate the jump in the Bayes factor seen in \fig{fig:single_psr_bf}. In \fig{fig:j0030_sims_bf} we show the results of the simulations and analyses. In each case the noise parameters obtained from the analysis of the real PSR J0030+0451 data were used to try and replicate the same Bayes factor. In the first four cases no GWB background was injected. Two different types of noise injection were used, corresponding to either the power law noise model values for this pulsar, or the free-spectral noise model parameters. When the same model is used in the analysis as the injection model the Bayes factor is near one. However, when a free-spectral noise model is used in the injection and a power law noise model is used in the analysis the Bayes factor roughly triples. This is further evidence in support of the main conclusion in this work \textemdash\ {\it the use of an inaccurate noise model can lead to anomalous detections of a GWB.}
%%%%%%% J0030+0451 Sims %%%%%%%%%%
%%%%%%%%%%%%%%%%%%%%%%%%%%%

One may question whether the free-spectral model will remove any evidence for a GWB entirely. The last two GWB Bayes factors in \fig{fig:j0030_sims_bf} show simulations where an actual power law GWB was injected in addition to the free spectral noise. The $A_{\rm GWB}$ injected was the maximum-likelihood value from the 7.5-year slice of PSR J0030+0451's real data set. One can see that the data analysis in the 7.5-year slice does not detect the GWB as a separate red noise injection. The full 11-year data set is able to better differentiate the GWB. This supports the conclusion from \sect{sec:fs_rn_model} and born out in the full time span analysis with free spectral models shown in \fig{fig:fs_ul}. One expects earlier slices to have lower Bayes factors, and a slow rise in the Bayes factor as we accrue longer data sets.

Obviously these Bayes factors are still rather close to one, i.e., even odds, but since the GWB signal significance is expected to grow slowly with time it is the comparison and trends with which we are most concerned.

\subsection{Power Accounting}
One would like to quantify the difference between the noise model using a power law prior for the Gaussian processes and the free spectral models that have effectively mitigated the \kink, removing the false-positive detection of a GWB. One way in which this can be done is to calculate the posterior probability distributions for the power in these red noise channels. Here we compare the power by calculating it from the models used as priors for the Gaussian process over the frequencies sampled in the Gaussian process coefficients. For the power law this will be a sum of the power law values across the sampled frequencies times the frequency bin size,
\begin{equation}\label{eqn:power_powerlaw}
	P = \sum^{N_f}_{i}A^2_{\rm RN}\left(\frac{f_i}{f_{yr}}\right)^{-\gamma}\;\mathrm{yr}^3 \Delta f .
\end{equation}

\begin{figure}[htp]\label{fig:power_j0030}

\includegraphics[clip,width=\columnwidth]{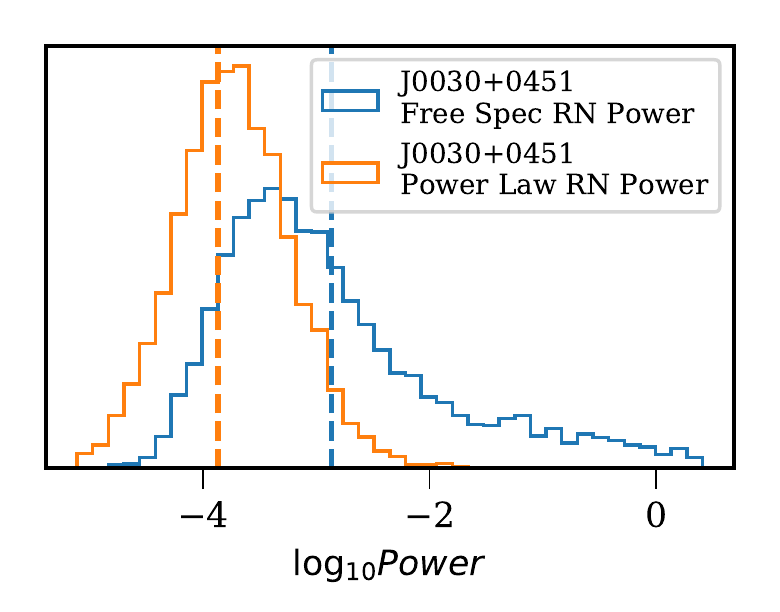}
 
\caption{Power posteriors for PSR J0030+0451 comparing the power in the power law model versus the free spectral model. The mean values for the power are different by roughly an order of magnitude. }

\end{figure}

In the case of the free spectral model the amplitudes are indexed, i.e.,
\begin{equation}\label{eqn:power_freespec}
	P = \sum^{N_f}_{i}\rho_i^2\;\mathrm{yr}^3 \Delta f .
\end{equation}
In \fig{fig:power_j0030} we show the calculated posteriors for the power using these two models on the data from PSR J0030+0451. The white noise parameter posteriors are basically unchanged for this pulsar between the two models, so the free-spectral model is effectively absorbing power that is otherwise unmodeled. While the lowest frequency is most likely the issue for this pulsar, as mentioned in \S~\ref{sec:results}, the majority of the power in the free-spectral model is at high frequencies, and is probably not to blame for the spurious GWB detection in the previous section. However, since the power posteriors for all of the culprit pulsars are different by approximately an order of magnitude between the power law and free spectral models, this noise has been flagged as an obvious area for improvement in our per pulsar noise modeling. A number of in-progress projects and a forthcoming paper, \citep{Simon:2019}, are devoted to mitigating noise of this sort in a number of NANOGrav pulsars.
%%%%%%% Power %%%%%%%%%%
%%%%%%%%%%%%%%%%%%%%%

%%%%%%% INEFFECTIVE ANALYSES %%%%%%%%%%%%%%
\subsection{Ineffective Analyses }\label{sec:misc}
While the use of the free-spectral model has mitigated the spurious GW signal of \fig{fig:std_ul}, there are a number of additional investigated analyses that either had a neutral or minimal effect on the GW statistics. Here we summarize them, in part to inform the interested expert and in part to motivate their use in upcoming work that investigates more comprehensive noise models for pulsars.

The standard analyses were run with various versions of the JPL solar system ephemeris and the Bayesian solar system ephemeris model, BayesEphem \citep{abb+17b}. All results were qualitatively the same as results shown in the sections above between DE421, DE430, DE436 \citep{2009IPNPR.178C...1F,2014IPNPR.196C...1F,de436} and using BayesEphem. In particular, all results showed the same anomalous GW statistics in the 6.5 to 8.5 year slices.

The choice of frequencies sampled by the Gaussian processes that are modeling the underlying stochastic signals has been shown to affect the signal analysis \citep{vanHaasteren:2014faa,Ellis:2016mtg}. A number of strategies for choosing the frequencies, including using a log-spacing, rather than a linear-spacing and choosing the frequencies uniformly across the slices, were carried out. These had some effect in mitigating the \kink\, but were not as effective as the methods described in previous sections. 

Lastly the proximity in time of the \kink\ to the changeover in the backends used for observing is intriguing, however, analyses testing any causation were inconclusive. These included  a number of analyses with different combinations of the overlapping GASP/ASP and GUPPI/PUPPI data. Here it is difficult to separate the effect of narrower bandwidths from the change in GW statistics. We also modeled backend red noise, similar to the ``band noise'' of  \citet{Lentati:2016ygu}, but rather than restricting the noise to a specific observing frequency band, we restricted the noise to a specific backend type. While the red noise parameters seem to be significantly different between backends in some pulsars, the use of these models did not help to mitigate the spurious GW statistics\footnote{Anyone interested in other analyses undertaken during this research, or seeing the results of those discussed in this subsection should feel free to contact the first author for more information.}. 

%%%%%%%%%%MITIGATING LOW SPECTRAL INDEX NOISE%%%%%%%%%%%%%%%
\section{Mitigating Low Spectral Index Noise}
\label{sec:lowspec}
Here we turn our attention to mitigating the transient WN feature described at the end of \sect{sec:initial_slice_results}.
The analysis in which the spectral index is varied intimates PSR J1713+0747 as a strong candidate for this WN event in the {\it common} red noise process. 
Sharp noise features in the time series domain can manifest as low-spectral-index noise in the power spectral density, hence an attempt at mitigating the noise in this pulsar was undertaken. 

The recent observation of a second ISM event in PSR J1713+0747 has prompted new work on chromatic noise models for our pulsars \citep{Lam:2017duo}. This second event does not occur during the time span of the NG 11-year dataset, but the first ISM event occurring near MJD $54750$ (2008.78) and first observed in \cite{dfg+13} does, between the 4.0 year and 4.5 year slices in this analysis. As has been shown in other recent publications \citep{Aggarwal:2018mgp, Aggarwal:2019b}, unmodeled noise in a single pulsar can appear in a common PTA signal. In order to investigate whether the first ISM event in PSR J1713+0747 is the cause of the significant white noise appearing in the common red noise signal we ran the analysis over again using the same DM noise model\footnote{This type of model has been used previously in other analyses \citep{Lentati:2016ygu}.} first used in \cite{Lam:2017duo}. That model consists of a timing-model fit for a linear and quadratic trend in the DM  (DM1 and DM2), chromatic red noise modeled with a Gaussian process and a phenomenological model for the dip in the DM variations. In \cite{Lam:2017duo} this consisted of two exponential dips modeled as 
\begin{equation}
    \Delta \mathrm{DM}_{\rm dip}=-\mathcal{T}\Theta\left(t_0\right)\exp \left(-(t-t_0)/\tau\right)
\end{equation}
where $\Theta\left(t_0\right)$ is the Heaviside function and the amplitude ($\mathcal{T}$), time of occurrence ($t_0$) and decay time ($\tau$) were fit for in the Bayesian analysis. In the present work we only fit for one exponential dip, to model the first ISM event. The Gaussian process and exponential dips are implemented in {\tt enterprise}.  This model replaces the piece-wise DMX model used in \citetalias{abb+17} and \citetalias{abb+17b}. This model was also studied in depth in \citet{Wang:2019xfx} where a Bayesian cross-validation study showed convincing evidence for the preference of this model. The better performance of this model for DM variations in this particular case is explained by a lack of DMX bins in the time span around the minimum of the rapid fluctuation. From MJD $54707$ (2008.66) until 100 days past the event there are only 5 DMX bins. This coarse sampling is possibly inadequate for such a relatively short-timescale, high-amplitude event. The lack of inter-observing band TOAs also limits the precision of the $\Delta$DM measurement, because the measurement is done within a single, narrow receiver band.

\begin{figure}[htp]\label{fig:vary_gamma2}

\includegraphics[clip,width=\columnwidth]{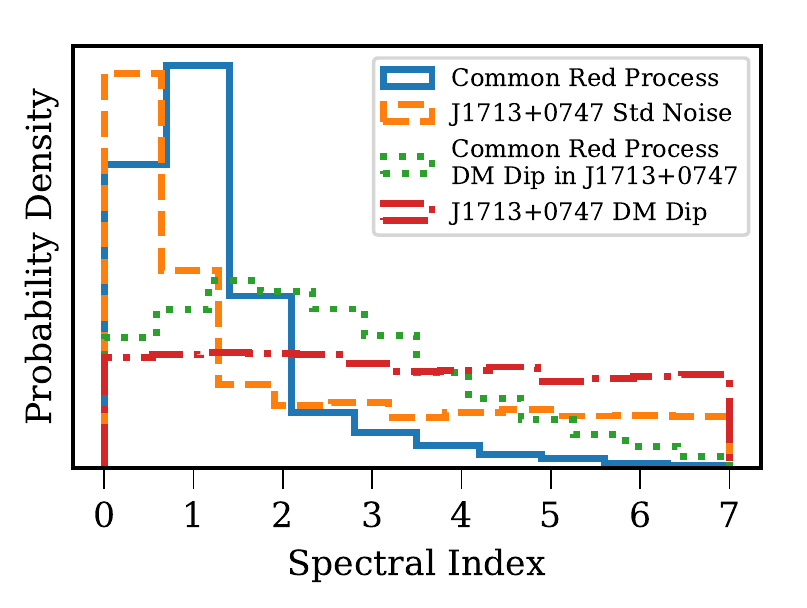}
 
\caption{Comparison of posteriors for the spectral index of a power law red noise process. Posteriors are shown for both the full PTA analysis and the individual PSR J1713+0747 noise analysis. Note that using a more tailored noise model on this one pulsar has a significant impact on the spectral index of the common process in the full PTA.}

\end{figure}

The results of this newer model on the posteriors for the spectral index are dramatic, as can be seen in \fig{fig:vary_gamma2}.
%%%%%%%% Compare Varying Gamma at 4.5 Year Slice %%%%%%%%%%
The posteriors show reduced significance of an unmodeled WN transient, i.e., the posterior is no longer butted up against the $\gamma=0$ end of the prior. Not only is this noise mitigated in the individual pulsar noise, but also in the common process, revealing that this event can have an important effect on the GWB analysis and can be mitigated with a more tailored noise model\footnote{It should be noted that there appears to be a large amount of power at lower spectral indices, shallower than expected from either pulsar spin noise or a GWB. This points to a systemic issue with the current noise models used in PTA gravitational-wave analysis and will be confronted in an upcoming NANOGrav publication \citep{Simon:2019}.}.

With the appearance of a second ISM event in PSR J1713+0747 in the NANOGrav 12.5-year data set this type of model will be necessary to properly use this pulsar in our GW analyses.

%%%%%%%%%%%SUMMARY AND CONCLUSIONS%%%%%%%%%%%%%%%%
\section{Summary and Conclusions}
\label{sec:conclusions}
Here we have used the standard tools of PTA gravitational-wave analysis to investigate the evolution of gravitational-wave statistics in the NANOGrav 11-year data set. 
After finding transient features in the sliced data GW-analyses we undertook an in-depth analysis to characterize and mitigate any possible sources of noise that might lead to these features. 
The transient GWB detection that peaks during the $7.5$-year slice was found to be due, in large part, to five ``culprit'' pulsars. 
These pulsars were identified by a combination of their individual red noise analyses, single pulsar GWB upper-limits and a new Bayesian PTA data analysis technique known as the dropout method. 
In order to test whether these pulsars were responsible for the large GWB signal, a new set of GW statistics were derived for the NANOGrav 11-year data set with TOAs from these pulsars removed. 
Once a set of pulsars was identified to be responsible for these artifacts the false signal was mitigated using a free spectral noise model. 
This demonstrates the importance of characterizing the noise in pulsars correctly, and demonstrates that incorrect noise models can lead to false positives in our Bayesian analyses.
These results highlight a number of strategies important when searching PTA data for a stochastic GWB:
\begin{itemize}
	\item Study the noise {\it evolution} of individual pulsars.
	\item Look at the evolution of single-pulsar UL and detection analyses to see {\it when} various signals become significant, and {\it how long} they remain significant.
	\item Attempt to use more tailored noise models for a given pulsar.
\end{itemize}
This last point is especially important for the type of transient WN feature seen in the $4.5$-year slice. As shown in \S\ref{sec:lowspec} this transient feature, causing the varying-spectral index analysis to prefer a spectral index of zero, and contemporaneous with the first interstellar medium event in PSR J1713+0747, was mitigated using a phenomenological model for the chromatic time delays consisting of a Gaussian process $+$ exponential dip. 

This work has shown that the standard tools for GW analysis in pulsar timing data, while sufficient to mitigate some of the noise features in \citetalias{abb+17}, need to be updated as the sensitivity of our detector has revealed a new noise floor. These considerations have moved NANOGrav to undertake a program of Bayesian model selection using a full suite of individual, tailored noise models for our pulsars that pay closer attention to the astrophysics causing the noise in each case. This will be presented in an upcoming paper \citep{Simon:2019} presenting these new models and the results of model selection on the most sensitive \citetalias{abb+17} pulsars. 

\acknowledgements

\emph{Author contributions.}This paper is the result of the work of dozens of people over the course of more than thirteen years. 
We list specific contributions below. JSH ran the sliced analyses and led the paper writing. JS, SRT, MTL, SJV, KI and JSK contributed substantially to paper writing, discussion and interpretation of results. MTL helped with analyses. JSH and SRT developed the formalism in \sect{sec:motivation}. ZA, KC, PBD, MED, TD, JAE, ECF, RDF, EF, PAG, GJ, MLJ, MTL, LL, DRL, RSL, MAM, CN, DJN, TTP, SMR, PSR, RS, IHS, KS, JKS, and WZ ran observations and developed the 11-year data set. 

\emph{Acknowledgments.}
The NANOGrav project receives support from National Science Foundation (NSF) PIRE program award number 0968296 and NSF Physics Frontier Center award number 1430284.
NANOGrav research at UBC is supported by an NSERC Discovery Grant and Discovery Accelerator Supplement and by the Canadian Institute for Advanced Research.
JSH would like to thank Joseph Romano for a number of in-depth discussions concerning the results of the numerous analyses done for this project.
MV and JS acknowledge support from the JPL RTD program.
Portions of this research were carried out at the Jet Propulsion Laboratory, California Institute of Technology, under a contract with the National Aeronautics and Space Administration.
SRT was partially supported by an appointment to the NASA Postdoctoral Program at the Jet Propulsion Laboratory, administered by Oak Ridge Associated Universities through a contract with NASA. SRT thanks ERS for fruitful discussions.
JAE was partially supported by NASA through Einstein Fellowship grants PF4-150120.
SBS was supported by NSF award \#1458952.
PTB acknowledges support from the West Virginia University Center for Gravitational Waves and Cosmology.
This work was supported in part by National Science Foundation Grant No.~PHYS-1066293 and by the hospitality of the Aspen Center for Physics.
Portions of this work performed at NRL are supported by the Chief of Naval Research.
This research was performed in part using the Zwicky computer cluster at Caltech supported by NSF under MRI-R2 award No.~PHY-0960291 and by the Sherman Fairchild Foundation.
A majority of the computational work was performed on the Nemo cluster at UWM supported by NSF grant No.~0923409.
Parts of the analysis in this work were carried out on the Nimrod cluster made available by S.M.R.
Data for this project were collected using the facilities of the National Radio Astronomy Observatory and the Arecibo Observatory.
The National Radio Astronomy Observatory and Green Bank Observatory are facilities of the NSF operated under cooperative agreement by Associated Universities, Inc.
The Arecibo Observatory is operated by the University of Central Florida, Ana G. M\'{e}ndez-Universidad Metropolitana, and Yang Enterprises under a cooperative agreement with the NSF (AST-1744119).
This research is part of the Blue Waters sustained-petascale computing project, which is supported by the National Science Foundation (awards OCI-0725070 and ACI-1238993) and the state of Illinois.
Blue Waters is a joint effort of the University of Illinois at Urbana-Champaign and its National Center for Supercomputing Applications.
Some of the algorithms used in this article were optimized using the Blue Waters allocation ``Accelerating the detection of gravitational waves with GPUs''.
The Flatiron Institute is supported by the Simons Foundation.

\bibliographystyle{aasjournal.bst}
\bibliography{bib}

\end{document}